\documentclass[fleqn,10pt]{wlscirep}
\pdfoutput=1
\usepackage[utf8]{inputenc}
\usepackage[T1]{fontenc}
\usepackage{adjustbox}
\usepackage{amsmath}
\usepackage{comment}

\usepackage{amsthm}
\usepackage{cite}
\usepackage{adjustbox}
\usepackage{caption}
\usepackage{mathtools}
\usepackage[multiple]{footmisc}
\usepackage{float}
\usepackage[caption = false]{subfig}
\usepackage{afterpage}
\usepackage{epstopdf}
\usepackage{tablefootnote}

\title{Temporal-topological properties of higher-order evolving networks}

\author[1,*]{Alberto Ceria}
\author[1]{Huijuan Wang}
\affil[1]{Faculty of Electrical Engineering,
Mathematics, and Computer
Science, Delft University of
Technology, Mekelweg 4, 2628 CD,
Delft, The Netherlands}
\affil[*]{A.Ceria@tudelft.nl}

\begin{abstract}
Human social interactions are typically recorded as time-specific dyadic interactions, and represented as evolving (temporal) networks, where links are activated/deactivated over time. However, individuals can interact in groups of more than two people. Such group interactions can be represented as higher-order events of an evolving network. Here, we propose methods to characterize the temporal-topological properties of higher-order events to compare networks and identify their (dis)similarities. 
We analyzed 8 real-world physical contact networks, finding the following: a) Events of different orders close in time tend to be also close in topology; b) Nodes participating in many different groups (events) of a given order tend to involve in many different groups (events) of another order; Thus, individuals tend to be consistently active or inactive in events across orders; c) Local events that are close in topology are correlated in time, supporting observation a). Differently, in 5 collaboration networks, observation a) is almost absent; Consistently, no evident temporal correlation of local events has been observed in collaboration networks. Such differences between the two classes of networks may be explained by the fact that physical contacts are proximity based, in contrast to collaboration networks.
Our methods may facilitate the investigation of how properties of higher-order events affect dynamic processes unfolding on them and possibly inspire the development of more refined models of higher-order time-varying networks.

\end{abstract}
\begin{document}

\flushbottom
\maketitle
%
%
\thispagestyle{empty}
\setlength{\parskip}{0pt}

\section{Introduction}

Interactions among individuals are usually experimentally measured as time-resolved records of face-to-face contacts between couples of people in controlled social setting such as workplaces, hospitals, schools and conferences. These time specific records are thus collected in the form of dyadic interactions, and have been effectively studied in the framework of evolving (temporal) networks, where each link between two nodes is activated only when the node pair interacts  \cite{holme2012temporal,holme2015modern,masuda2016guide}. The temporal patterns of link activations (or contacts) in real-world networks are far from being fully random nor deterministic \cite{karsai2018bursty}. Contacts between a pair of nodes usually occur in bursts of many contacts close in time followed by a long period of inactivity \cite{karsai2012universal} and the time between two consecutive interactions is usually fat-tailed distributed \cite{goh2008burstiness,oliveira2005darwin,eckmann2004entropy}. Such temporal properties of contacts influence the dynamic processes unfolding on the network  \cite{zhan2019information,zhan2019suppressing,miritello2011dynamical,horvath2014spreading,backlund2014effects,williams2019auto,karsai2011small,delvenne2015diffusion,unicomb2021dynamics}. Despite these tremendous advances in the last decade, studies on temporal networks have traditionally focused on pairwise interactions only. However pairwise interactions can only partially capture interactions among constituents of a system \cite{battiston2020networks,battiston2021physics}. For example, a neuron may receive the output from or send a signal to many different neighbouring neurons \cite{petri2014homological}, individuals may gather in groups \cite{sekara2016fundamental}, and scientific collaborations are not limited to couples of authors \cite{patania2017shape}. Such interactions are named higher-order, to emphasize that they involve more than just a couple of nodes.
Benson et al.\cite{benson2018simplicial} showed that a generalization of triadic closure seems to lead the first activation of a given hyperlink. On the other hand, Cencetti et al. \cite{cencetti2021temporal} focused on temporal inhomogeneities of activations of the same hyperlink. The focus so far is on the prediction of hyperlink activations \cite{benson2018simplicial} or on pure temporal properties of higher-order events \cite{cencetti2021temporal}. However, the interplay between temporal and topological properties of higher-order events, e.g. if higher-order events close in time tend to occur also close in topology, remains far from well understood. 
Hence, this work aims to systematically characterize the relation between temporal and topological properties of higher-order events to compare higher-order temporal networks. Inspired by our recent work that characterizes temporal and topological properties of dyadic interactions in temporal networks \cite{ceria2021topological}, we redesign the characterization method for higher-order events. In particular, we are going to explore such properties from three perspectives: 1) The interrelation between the distance in topology and the temporal delay of events, 2) Their correlation or overlap in topological location 3) The temporal correlation of local events that overlap in component nodes. In order to compare real-world networks with different sizes, we design null models where temporal and topological properties of events of an arbitrary order are systematically destroyed or preserved. We applied our methods to 8 real-world physical contact networks and 5 collaboration networks. We show that, in physical contacts, events of different orders with short temporal delay tend to be close in topology too. 
We then investigate the correlation of events in topology and discover that events of different orders are likely to overlap in component nodes.
In particular, nodes who participate in many different groups (events) of a given order are likely to be involved in many different groups (events) of another order. Individuals do not reduce their number of interactions of one order due to frequent interactions of another order.
Finally, we show that those local events that overlap in component nodes are correlated in time, which supports the finding that events close in time are also close in topology.
In collaboration networks, we observe that events also overlap in component nodes. However, the correlation between topological distance and temporal delay of events are usually either weak or absent. Coherently, in collaboration networks, the temporal correlation of local events that overlap in component nodes is almost absent. Such differences between physical contacts and collaboration networks may be due to the fact that physical interactions are partly driven by proximity, so that a set of individuals close to each other tend to interact close in time among (subsets of) them.

Our methods can be applied to compare real-world higher-order networks and to investigate how the properties of their events affects the dynamic processes unfolding on them.  More realistic models of higher-order evolving networks can be further developed to reproduce specific properties of the higher-order interactions observed in this paper.

\section{Definitions}
\subsection{Higher-order evolving networks}
Time-varying social interactions or contacts have been mostly measured pairwise and studied with the formalism of (pairwise) temporal networks. A temporal network observed at discrete time within $[0,T)$ can be described by $\mathcal{G} = (\mathcal{N}, \mathcal{C})$, where $\mathcal{N}$ is the set of nodes or individuals, $\mathcal{C}$ is the set of pairwise interactions. If node $u$ and $v$ have a contact at time step $0\leq t \leq T-1$, $(\ell,t) \in \mathcal{C}$, where $\ell = \ell(u,v)$ is the link connecting the pair of nodes between which the contact occurs. The contact $(\ell(u,v),t)$ can be regarded as the activation of the link $\ell(u,v)$ at time $t$. 
This traditional temporal network representation records social contacts as a set of pair-wise interactions. However, individuals may gather in larger groups, so that more than two people interact with each other at the same time. For example, an interaction $(h(i,j,k),t)$ among three nodes at time $t$ is usually measured and recorded as three pair-wise interactions $(\ell(i,j),t)$, $(\ell(j,k),t)$ and $(\ell(i,k),t)$. Social interactions can be more precisely represented as a higher-order evolving network $\mathcal{H} = (\mathcal{N},\mathcal{E})$ (or temporal hypergraph, following the definition of Cencetti et al. \cite{cencetti2021temporal}),
 where $\mathcal{E}$ is the set of events of arbitrary orders. Such group interaction or higher-order event $(h(u_1,\dots u_d),t)$ can be regarded as the activation of the corresponding hyperlink $h(u_1,\dots u_d)$ at $t$. The size or order of the interaction is $d$, where $d$ is the size of the group.  
 The pairwise time aggregated network of a traditional pairwise temporal network is $G = (\mathcal{N},\Lambda)$, where any couple of nodes $(i,j)$ is connected by a link $\ell(i,j) \in \Lambda$ if $\ell(i,j)$ has been active at least once during the entire observation time $[0,T)$. Consistently, the higher-order time aggregated network is $H = (\mathcal{N},\mathcal{L})$, where any set $\{u_1,\dots u_d\}$ of $d$ nodes are connected by a hyperlink $h(u_1,\dots u_d) \in \mathcal{L}$ with size $d$ if $h(u_1,\dots u_d)$ has been activated at least once. The activity of each hyperlink $h$ can be represented by a time series $X_{h} = \{x_{h}(t), 0\leq t < T\}$ where $x_{h}(t) = 1$ only if the hyperlink $h$ is active at time t, i.e., $e=(h,t) \in \mathcal{E}$. 
\subsection{Temporal and topological distance of events}
\label{sec:top_temp_distance}
The temporal distance or delay between two events $e_1 = (h_1, t)$ and $e_2 = (h_2, s)$ is $\mathcal{T}(e_1, e_2)=|t-s|$.

The topological distance, also called hop-count, between two nodes on a pair-wise static network is the number
of links contained in the shortest path between these two nodes. We define the topological distance
$\eta(e_1,e_2)$ between two events $e_1 = (h_1, t)$ and $e_2 = (h_2, s)$ as the topological distance between the corresponding two hyperlinks $h_1$ and $h_2$, which is further defined as follows. The distance between the same hyperlink is zero, e.g., $\eta((h_1,t),(h_1,s)) = 0$. The distance between two different hyperlinks $h(u_1,\dots,u_d)$ and $h(v_1,\dots,v_{d'})$ with size $d$ and $d'$, respectively, follows
\begin{equation}
    \eta((h(u_1,\dots,u_d),t),(h(v_1,\dots,v_{d'}),s)) = min_{u\in \{ u_1,\dots,u_d\}, v \in \{v_1,\dots,v_{d'}\}}(\delta(u,v)+1)
\label{eq:eq1}
\end{equation}
where $\delta(u,v)$ is the distance or hop-count between node $u$ and $v$ on the unweighted pairwise time aggregated network $G$. The distance between two events is thus one plus the minimal distance between two component nodes from the two events respectively. For example, the distance between events $ e_1 = (h(i,j,k),t)$ and $e_2 = (h(i,m,n),s)$ is $\eta(e_1,e_2) = 1$. 
\subsection{Network randomization - control methods}
To detect non-trivial temporal and topological patterns of events, we compare properties obtained from real-world higher-order temporal networks with those of designed null models. We generalize the randomized reference models of pairwise evolving networks which gradually preserve and destroy temporal and topological properties of pairwise interactions \cite{gauvin2018randomized,nakajima2021randomizing,ceria2021topological} for higher-order temporal networks.  Given a higher-order evolving network $\mathcal{H}$ and any given order $d$ of events, we introduce 3 randomized null models $\mathcal{H}^1_d$, $\mathcal{H}^2_d$ and $\mathcal{H}^3_d$ which systematically remove or preserve specific temporal or topological properties of order $d$ events only, while preserving the properties of events of any other size $d'\neq d$. We denote as $\mathcal{E}_d$ the set of events with the same size $d$. Randomized network $\mathcal{H}^1_{d}$ is obtained by randomly re-shuffling the time stamps of the events in $\mathcal{E}_d$, without changing the topological locations of these events. This randomization does not change the total number of activations of each hyperlink, nor the probability distribution of the topological distance of two randomly selected events. As mentioned in Subsection \ref{sec:top_temp_distance}, the activations of a given hyperlink $h$ can be represented by a time series $X_{h}$. The randomized network $\mathcal{H}^2_d$ is obtained by iteratively swapping the time series of two randomly selected hyperlinks of size $d$. In $\mathcal{H}^2_d$, the inter-event time distribution of the activity of a random hyperlink of order $d$ is preserved as in the original network $\mathcal{H}$. The third randomized network $\mathcal{H}^3_d$ is obtained
by swapping the activity time series of two randomly selected hyperlinks with the same size $d$ and the same total number of activations. This randomization does not change the number of activations of any hyperlink, the distribution of the topological distance of two random events, nor the inter-event (order $d$ events) time distribution. These three randomized models preserve the unweighted higher-order time aggregated network $H$ and the probability distribution of the temporal distance of two random events of size $d$. 
\section{Datasets}
We will apply our method to 13 real-world datasets of human physical interactions and scientific collaborations. 
The first 8 datasets are collections of face-to-face interactions at a distance smaller than 2 m in several social contexts such as conferences (HT2009, SFHH), hospital, primary school (PS), high schools (HS2012,HS2013), workplace (WP2) and museum (Infectious).
Face-to-face interactions are recorded as a set of pair-wise interactions. Based on them, we deduce group interactions, by promoting each set of $\binom{d}{2}$ dyadic interactions occurring at the same time and forming a fully connected clique of $d$ nodes to an event of size $d$. Since a clique of order $d$ contains all its sub-cliques of order $d'<d$, only the maximal clique is promoted to a higher-order event, whereas sub-cliques are ignored. For example, 3 pairwise contacts $(\ell(i,j),t),\ (\ell(j,k),t)$ and $(\ell(i,k),t)$ occurring at the same time $t$ are regarded as a single event of order 3 i.e., $(h(i,j,k),t)$ without any order 2 event. This method has been already used by Cencetti et al \cite{cencetti2021temporal}. to deduce higher-order interactions from datasets of human face-to-face interactions. We further preprocess these datasets by removing nodes which are not connected to the largest connected component in the pairwise time-aggregated network. We also remove long periods of inactivity, when no event occurs in the network. Such periods usually correspond, e.g., to night and weekends, and are recognized as outliers in the inter-event time distribution of the time series which records the total number of events per timestamp. Such data pre-processing method has also been used in our recent work\cite{ceria2021topological}.
The other 5 higher-order collaborations networks are obtained based on scientific papers recorded in the arxiv in various fields: lattice high energy physics (hep-lat), theoretical nuclear physics (nucl-th), quantitative biology (q-bio), quantitative finance (q-fin) and quantum physics (quant-ph). In a collaboration network, each node represents an author, and an event of order $d$ occurrs at time t if a paper co-authored by $d$ authors is published at t.
Assigning papers to the correct authors is not easy. The same author can be named differently, e.g., using the full or initial of the first name and typographic errors may be present. Thus, we applied standard text preprocessing methods
to authors' name, and we identify each author by the initials of their first names, together with their surname according to the method of Newman et al.\cite{newman2001structure}.  
The total number of events of each order in each real-world temporal network is shown in Appendix (Figures  \ref{fig:hypercount_phys} and \ref{fig:hypercount_collab} in Supplementary Material): In each dataset, the number of events with order $2\leq d \leq 4$ is not negligible; however events with an order larger than 4 are rare (if not absent) in most of the physical contact datasets. 
Details of the datasets after preprocessing are given in Table \ref{tab:1}.

\begin{table}[H]
\centering
\begin{adjustbox}{width=0.8\columnwidth}
\begin{tabular}{@{}l|l|l|l|l|l|l|l|r@{}}
\toprule Network     & $|\mathcal{N}|$  &  $|\mathcal{L}|$ & $|\mathcal{E}|$  & $T$& $dt$  
&\textit{contact type}   \\ 
\hline
\midrule Primary School (PS)     &    242     &    12704     &    106877     &    3099     &    20 s     &    \textit{physical}       \\

High School 2013 (HS2013)     &    327     &    7818     &    172031     &    7371     &    20 s     &    \textit{physical}       \\

Hypertext 2009 (HT2009)     &    113     &    2434     &    19037     &    7227     &    20 s     &    \textit{physical}       \\

Infectious (Infectious)     &    410     &    3350     &    14275     &    1422     &    20 s     &    \textit{physical}       \\

Workplace 2015 (WP2)     &    217     &    4909     &    73820     &    20947     &    20 s     &    \textit{physical}       \\

SFHH Conference (SFHH)     &    403     &    10541     &    54306     &    3800     &    20 s     &    \textit{physical}       \\

Hospital (Hospital)     &    75     &    1825     &    27835     &    16027     &    20 s     &    \textit{physical}       \\

High School 2012 (HS2012)     &    180     &    2645     &    42105     &    14115     &    20 s     &    \textit{physical}       \\

High energy physics, lattice (hep-lat)     &    10598     &    11588     &    18267     &    10809     &    1 d     &    \textit{collaboration}       \\

Nuclear physics, theory (nucl-th)     &    25246     &    27094     &    39511     &    10620     &    1 d     &    \textit{collaboration}       \\

Quantitative biology (q-bio)     &    45645     &    22978     &    25973     &    10704     &    1 d     &    \textit{collaboration}       \\

Quantitative finance (q-fin)     &    7509     &    6192     &    7577     &    9027     &    1 d     &    \textit{collaboration}       \\

Quantum physics (quant-ph)     &    56036     &    70119     &    88769     &    10600     &    1 d     &    \textit{collaboration}       \\

\end{tabular}
\end{adjustbox}
\caption{Basic features of the empirical higher-order time-evolving networks after data processing. The number of nodes ($|\mathcal{N}|$), the number of hyperlinks ($|\mathcal{L}|$), the total number of events ($|\mathcal{E}|$), the length of the observation time window in time steps ($T$), the time resolution or duration of each time step ($dt$) in seconds or days and the contact type are shown.}
\label{tab:1}
\end{table}

\section{Characterizing temporal-topological properties of networks}
In this Section we introduce a systematic characterization method of higher-order temporal networks. We characterize the temporal and topological properties of events from three different perspectives. In Subsection \ref{subsec:topo_tempo_interr}, we analyze the interrelation between the temporal and topological distance of two arbitrary events of  different orders. In Subsection \ref{subsec:topo_corr}, we study the topological correlation of events, i.e., how events of different orders overlap in component nodes. Finally, Subsection \ref{subsec:loc_temp} introduces a method to characterize the temporal correlation of events occurring close in topology. 

\subsection{Correlation of temporal and topological distance of events}
\label{subsec:topo_tempo_interr}

\begin{figure}[H]
    \centering
    \includegraphics[width = \textwidth]{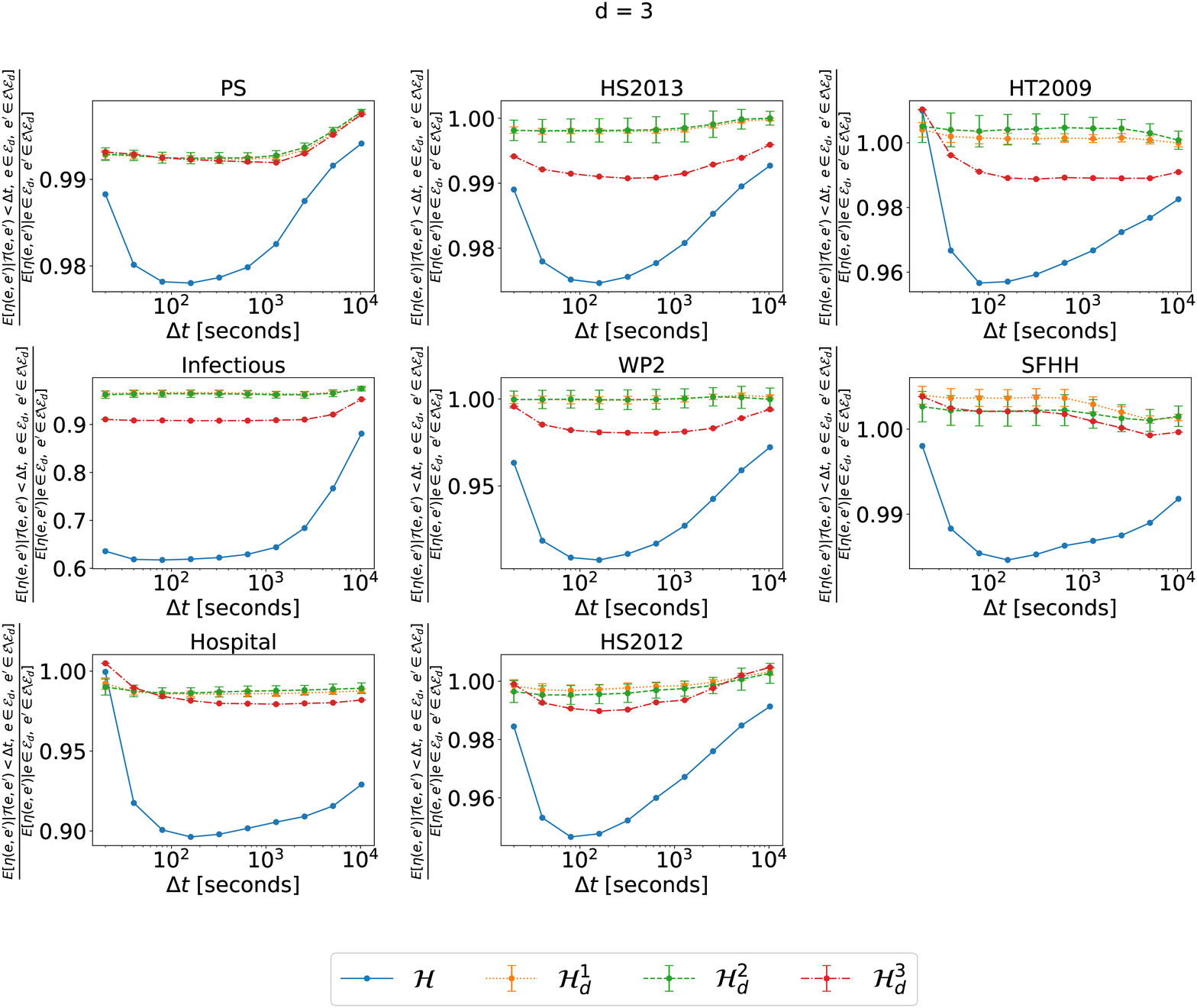}
    \caption{The normalized average topological distance $\frac{E[\eta(e,e') | \mathcal{T} (e,e') < \Delta t,\ e\in \mathcal{E}_d,\ e' \in \mathcal{E}\setminus \mathcal{E}_d ]}{E[\eta(e,e')|\ e\in \mathcal{E}_d,\ e' \in \mathcal{E}\setminus \mathcal{E}_d]}$, between an order $d=3$ event and an event of a different order, in each physical contact network and its corresponding three randomized null models $\mathcal{H}^1_d$ (yellow), $\mathcal{H}^2_d$ (green) and $\mathcal{H}^3_d$ (red), which preserve or destroy specific properties of order $d=3$ events. $\lim_{\Delta t\to\infty} E[\eta(e,e') | \mathcal{T} (e,e') < \Delta t,\ e\in \mathcal{E}_d,\ e' \in \mathcal{E}\setminus \mathcal{E}_d ] =E[\eta(e,e')|\ e\in \mathcal{E}_d,\ e' \in \mathcal{E}\setminus \mathcal{E}_d]$ for any $d$. The horizontal axes are presented in logarithmic scale.
    For each dataset, the results of the three corresponding randomized models are obtained from 10 independent realizations.}
    \label{fig:temp_top_cross}
\end{figure}
\begin{figure}[H]
    \centering
    \includegraphics[width = \textwidth]{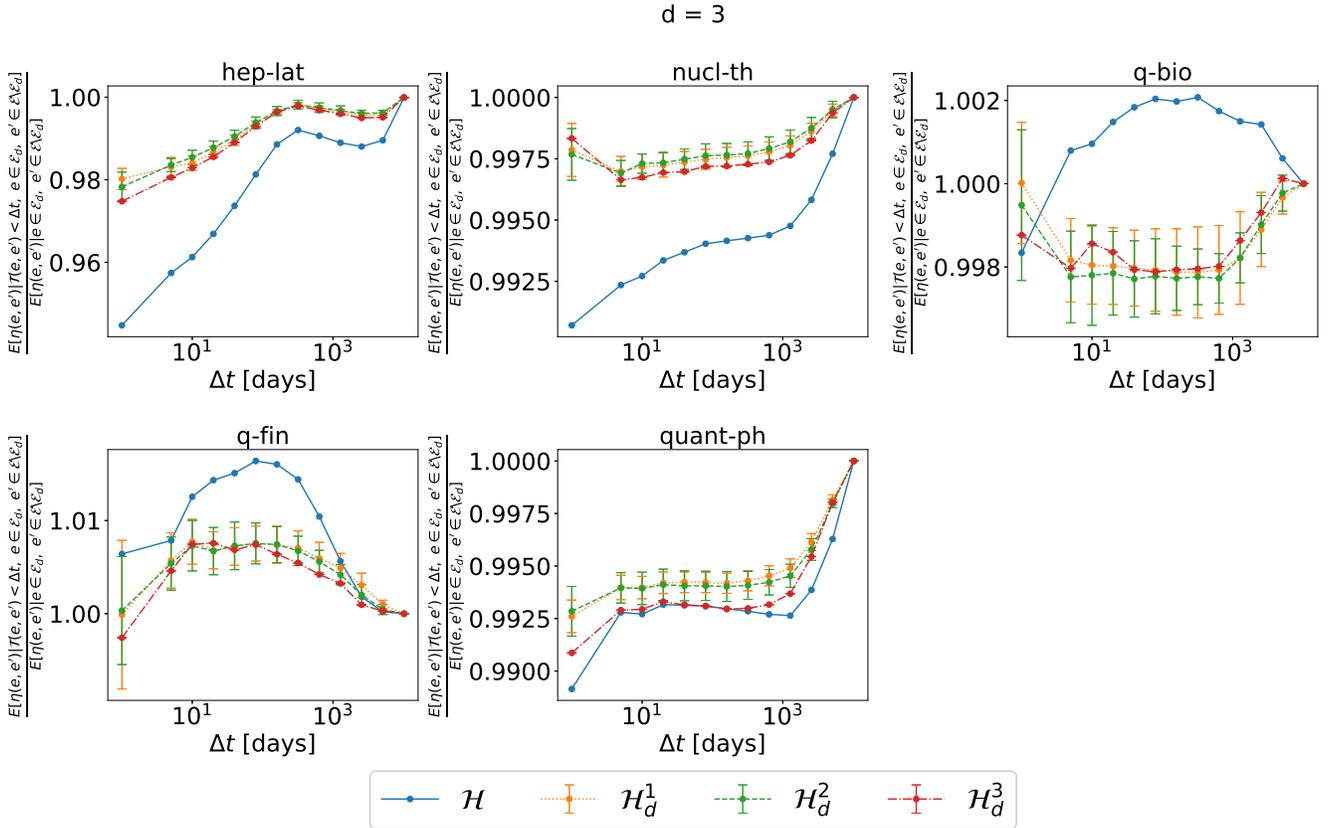}
    \caption{The normalized average topological distance $\frac{E[\eta(e,e') | \mathcal{T} (e,e') < \Delta t,\ e\in \mathcal{E}_d,\ e' \in \mathcal{E}\setminus \mathcal{E}_d ]}{E[\eta(e,e')|\ e\in \mathcal{E}_d,\ e' \in \mathcal{E}\setminus \mathcal{E}_d]}$, between an order $d=3$ event and an event of a different order, in each collaboration network and its corresponding three randomized null models $\mathcal{H}^1_d$ (yellow), $\mathcal{H}^2_d$ (green) and $\mathcal{H}^3_d$ (red), which preserve or destroy specific properties of order $d=3$ events. $\lim_{\Delta t\to\infty} E[\eta(e,e') | \mathcal{T} (e,e') < \Delta t,\ e\in \mathcal{E}_d,\ e' \in \mathcal{E}\setminus \mathcal{E}_d ] =E[\eta(e,e')|\ e\in \mathcal{E}_d,\ e' \in \mathcal{E}\setminus \mathcal{E}_d]$ for any $d$. The horizontal axes are presented in logarithmic scale.
    For each dataset, the results of the three corresponding randomized models are obtained from 10 independent realizations.}
    \label{fig:temp_top_cross_coll}
\end{figure}

In this subsection we investigate how temporal and topological distance of events are related to each other. Specifically, we aim to understand to what extent events close in time are also close in topology. In our previous work \cite{ceria2021topological}, we considered all interactions in a temporal network as pairwise interactions alone and found in real-world physical and virtual contact networks that pairwise interactions that are close in time tend to be close in topology (in the pairwise time aggregated network).
Here, we generalize the method of characterizing the relation between topological and temporal distance of two dyadic interactions to that of two higher-order events with different orders. In this analysis, normalizations in topological distance and randomizations in networks have been applied so that we can compare real-world temporal networks with different properties in e.g., the number of nodes and contacts. 
We take order $d= 3$ as an example to illustrate our method and observations. 
In Figures \ref{fig:temp_top_cross} and \ref{fig:temp_top_cross_coll} we investigate the average topological distance $E[\eta[(e,e')|\mathcal{T}(e,e')<\Delta t,e \in \mathcal{E}_d,\ e' \in \mathcal{E}\setminus \mathcal{E}_d]$ between two events $(e,e') $ with different orders $d\neq d'$, given that their temporal distance is smaller than $\Delta t$ in physical contact and collaboration networks, respectively. 

We observe an increasing trend of the normalized average topological distance between events with their conditional temporal distance in physical contact networks, but generally not in collaboration networks. Thus, in physical contacts, events of different orders that occur close in time tend to be also close in topology. The slope of this increase indicates the relative strength of temporal-topological correlation. The highest slopes are observed in Infectious, Workplace and Hospital networks. In contrast, this slope is small around zero in the corresponding randomized network $\mathcal{H}_d^2$, $\mathcal{H}_d^2$ and $\mathcal{H}_d^3$. Hence, the randomization remove the temporal and topological correlation. 
Conclusions drawn from the discussion of Figures  \ref{fig:temp_top_cross} and \ref{fig:temp_top_cross_coll} hold for the other orders $d = 2$ (see Figures \ref{fig:temp_top2_cross} and \ref{fig:temp_top2_cross_collab} in Supplementary Material) and $d=4$ (see Figures \ref{fig:temp_top3_cross} and \ref{fig:temp_top3_cross_collab} in Supplementary Material).

\subsection{Topological correlation of events with different orders}
\label{subsec:topo_corr}
\begin{figure}[H]
    \centering
    \includegraphics[width = 0.95\textwidth]{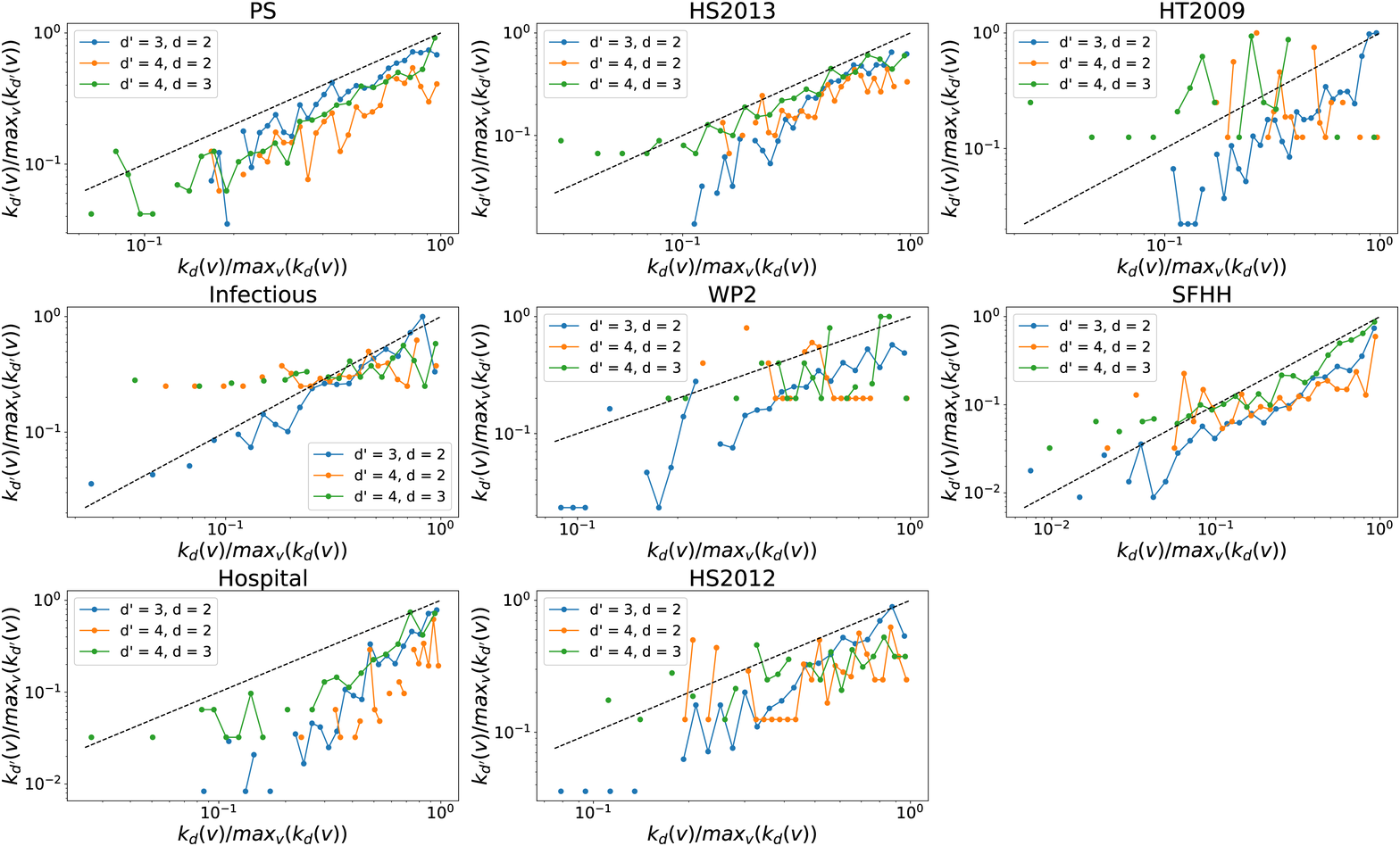}
    \caption{The $d'$-degree $k_{d'}(v)$ versus the the $d$-degree $k_{d}(v)$ of a node $v$ when $(d',d)$ is equal to (3,2) (blue line), (4,2) (yellow line) and (4,3) (green line) respectively in each physical contact network. Each axis (e.g., $k_{d}(v)$) has been normalized by its maximum (e.g., $max_{v}(k_{d}(v))$). 
    Only nodes whose $d$-degree and $d'$-degree are both non-zero are considered.
    The dashed line represent the reference case $\frac{k_{d'}(v)}{max_{v}(k_{d'}(v))} = \frac{k_{d}(v)}{max_{v}(k_{d}(v))}$. Note that both axes are presented in logarithmic scales. In total 30 logarithmic bins are split for horizontal axis.}
    \label{fig:node_corr_phys}
\end{figure}

\begin{figure}[H]
    \centering
    \includegraphics[width = 0.95\textwidth]{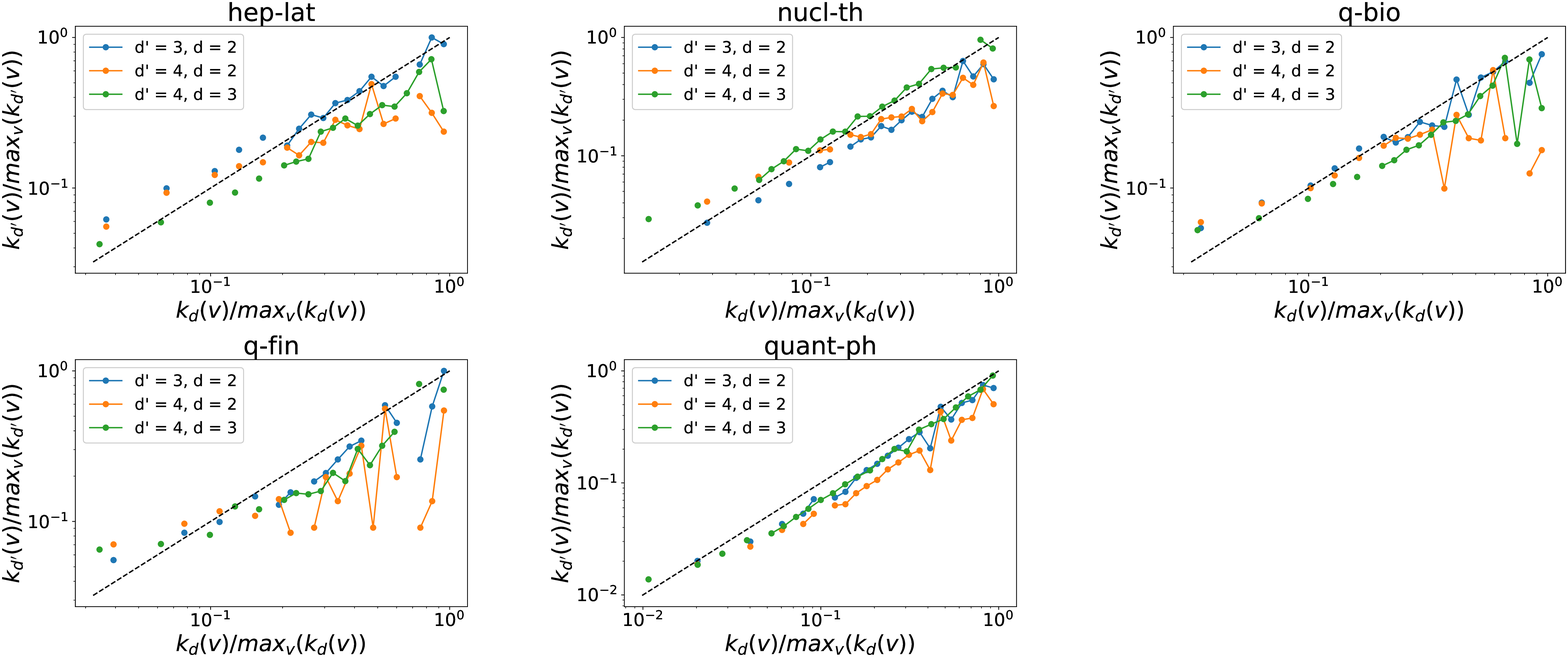}
    \caption{The $d'$-degree $k_{d'}(v)$ versus the the $d$-degree $k_{d}(v)$ of a node $v$ when $(d',d)$ is equal to (3,2) (blue line), (4,2) (yellow line) and (4,3) (green line) respectively in each collaboration network. Each axis (e.g., $k_{d}(v)$) has been normalized by its maximum (e.g., $max_{v}(k_{d}(v))$). 
    Only nodes whose $d$-degree and $d'$-degree are both non-zero are considered.
    The dashed line represent the reference case $\frac{k_{d'}(v)}{max_{v}(k_{d'}(v))} = \frac{k_{d}(v)}{max_{v}(k_{d}(v))}$. Note that both axes are presented in logarithmic scales. In total 30 logarithmic bins are split for horizontal axis.}
    \label{fig:node_corr_collab}
\end{figure}

To better understand the observed correlation between temporal and topological distance of events, we explore further whether higher-order events overlap in component nodes (correlation in topology) in this subsection and whether events that overlap in topology are correlated in time in Subsection \ref{subsec:loc_temp}. Higher-order events that overlap in component nodes and occur close in time may partially explain the observed temporal and topological correlation between events. 
Would a node that belongs to many hyperlinks of order $d$, also be connected to many hyperlinks of order $d'\neq d$? To investigate this question, we examine the number of hyperlinks of each order that a node belongs to in the higher-order time aggregated network.
The total number of order $d$ hyperlinks that the node $u$ is connected to, denoted as $k_d(v)$, is also called the $d$-degree of node $v$.
In Figure \ref{fig:node_corr_phys} (\ref{fig:node_corr_collab}), we compare the $d$-degree and the $d'$-degree of a node when $(d',d)$ is equal to (3,2), (4,2) and (4,3) respectively in each physical contact (collaboration) network. 
We focus on the case when $(d',d)$ is equal to (3,2), as an example. We observe that the $d'$-degree of a node is an increasing function of the $d$-degree of the node in every considered collaboration and physical contact networks. Hence, a node that participates in many groups of order $3$, tends to involve in many groups of order $2$. When $(d',d)$ equals to (4,2) and (4,3), such trend is less evident in physical networks (especially in WP2, HS2012, Infectious and HT2009) and remains evident in collaboration networks. This is likely because the number of order 4 hyperlinks is generally low (see Figure \ref{fig:active_hyper} in Supplementary Material) in physical contact networks, but not in collaboration networks (see Figure \ref{fig:active_hyper_collab} in Supplementary Material).

\begin{figure}[H]
    \centering
    \includegraphics[width = 0.95\textwidth]{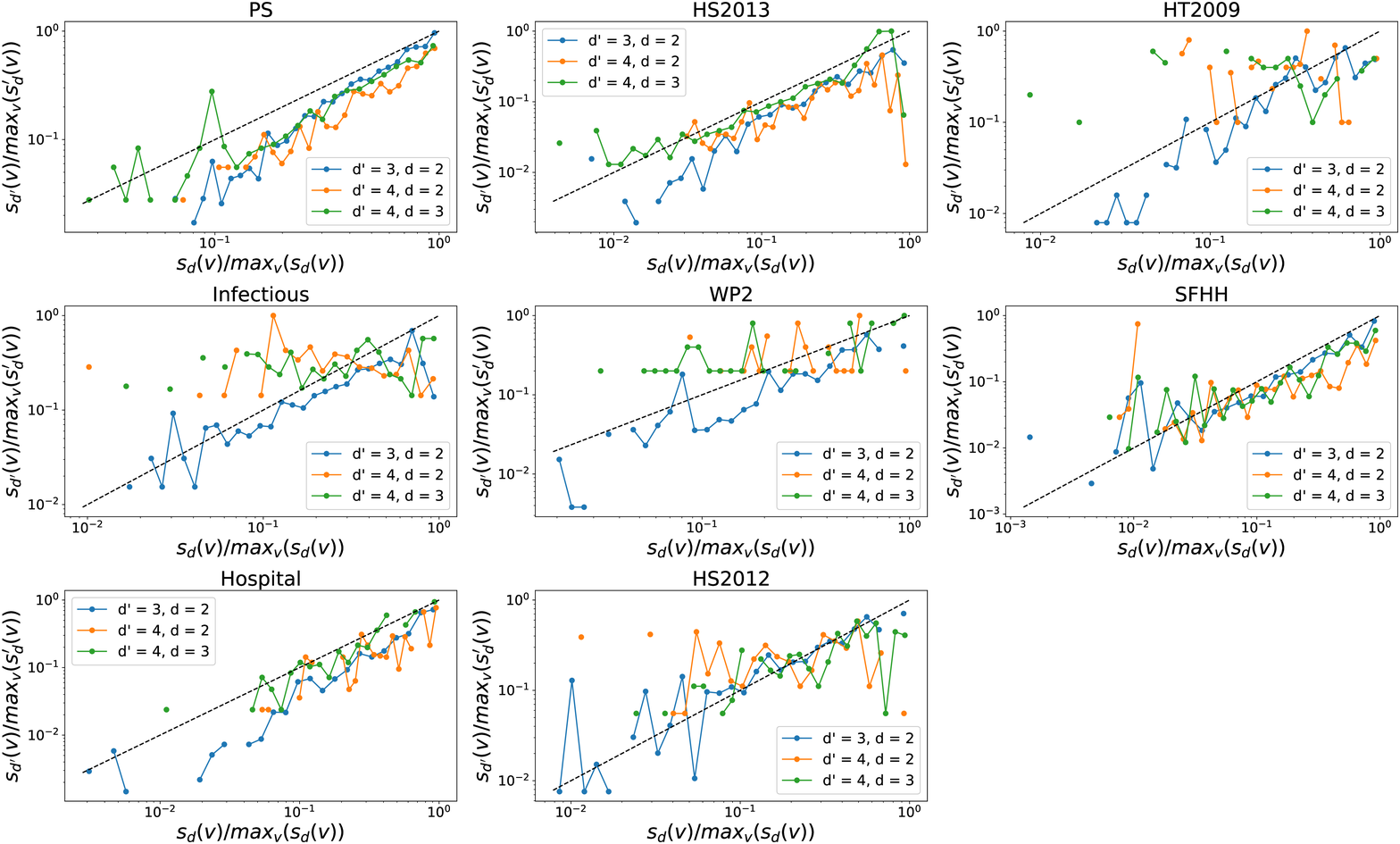}
    \caption{The $d'$-strength $s_{d'}(v)$ versus the the $d$-strength $s_{d}(v)$ of a node $v$ when $(d',d)$ is equal to (3,2) (blue line), (4,2) (yellow line) and (4,3) (green line) respectively in each physical contact network. Each axis (e.g., $s_{d}(v)$) has been normalized by its maximum (e.g., $max_{v}(s_{d}(v))$). Only nodes whose $d$-strength and $d'$-strength are both non-zero are considered.
    The dashed line represent the reference case $\frac{s_{d'}(v)}{max_{v}(s_{d'}(v))} = \frac{s_{d}(v)}{max_{v}(s_{d}(v))}$,  where $d'$-strength is a linear function of the $d$-strength of nodes. Note that both axes are presented in logarithmic scales. In total 30 logarithmic bins are split for horizontal axis.}
    \label{fig:node_str_corr_phys}
\end{figure}

\begin{figure}[H]
    \centering
    \includegraphics[width = 0.95\textwidth]{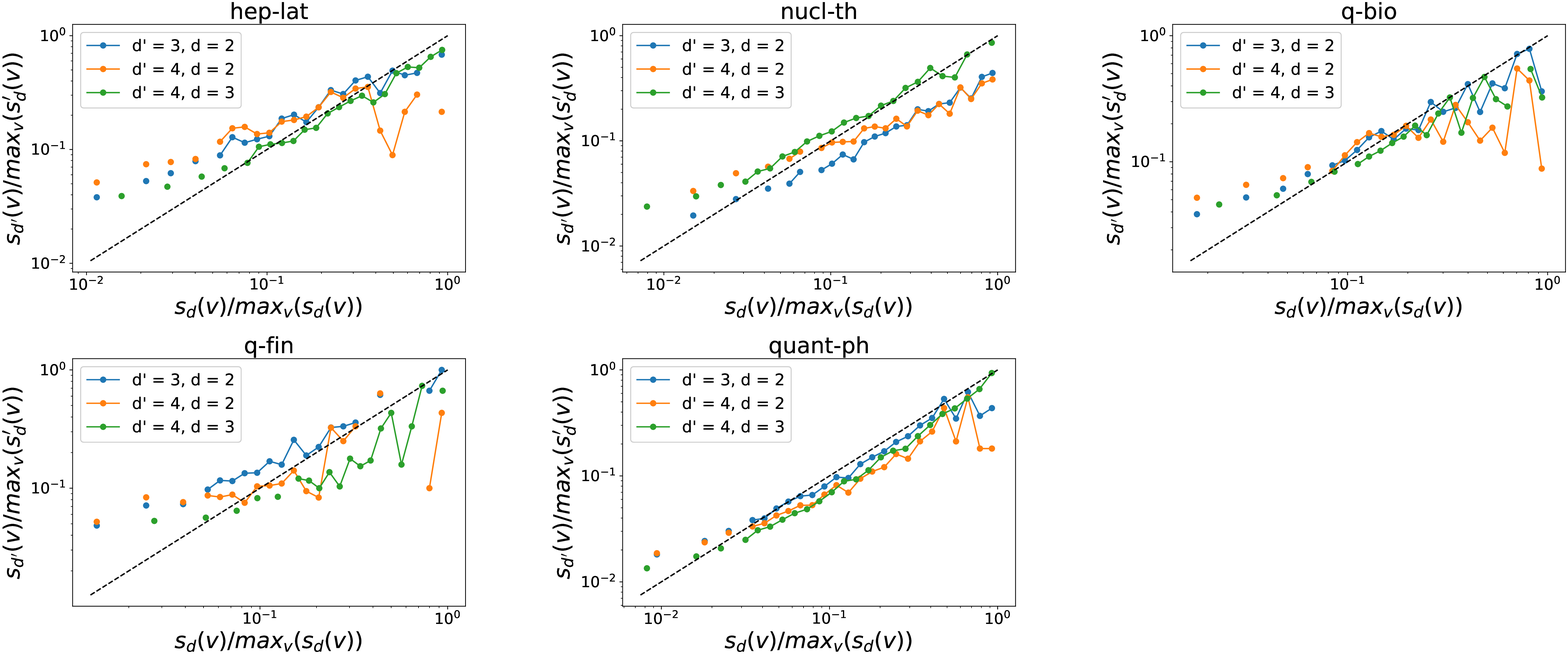}
    \caption{The $d'$-strength $s_{d'}(v)$ versus the the $d$-strength $s_{d}(v)$ of a node $v$ when $(d',d)$ is equal to (3,2) (blue line), (4,2) (yellow line) and (4,3) (green line) respectively in each collaboration network. Each axis (e.g., $s_{d}(v)$) has been normalized by its maximum (e.g., $max_{v}(s_{d}(v))$). 
    Only nodes whose $d$-strength and $d'$-strength are both non-zero are considered. The dashed line represent the reference case $\frac{s_{d'}(v)}{max_{v}(s_{d'}(v))} = \frac{s_{d}(v)}{max_{v}(s_{d}(v))}$,  where $d'$-strength is a linear function of the $d$-strength of nodes. Note that both axes are presented in logarithmic scales. In total 30 logarithmic bins are split for horizontal axis.}
    \label{fig:node_str_corr_collab}
\end{figure}

Furthermore, we investigate whether a node that involves in many order $d$ events tends to join many order $d'$ interactions. The number of order $d$ events that a node $v$ is involved in, denoted by $s_d(v)$, is also called the $d$-strength of node $v$. Similar to our analysis of the $d$-degree and $d'$-degree of node, we find the $d$-strength and $d'$-strength of a node are also positively correlated when $(d',d)$ equal to (3,2) in each temporal network, as shown in Figures \ref{fig:node_str_corr_phys} and \ref{fig:node_str_corr_collab}. This trend is less evident only in physical contacts that have few order $4$ events, when $(d',d)$ is equal to (4,3) and (4,2). This suggests that an individual's large number of interactions of one order would not reduce his or her number of events of another order. Individuals tend to be consistently active or inactive in events across orders. 
\begin{figure}[H]
    \centering
    \includegraphics[width = 0.95\textwidth]{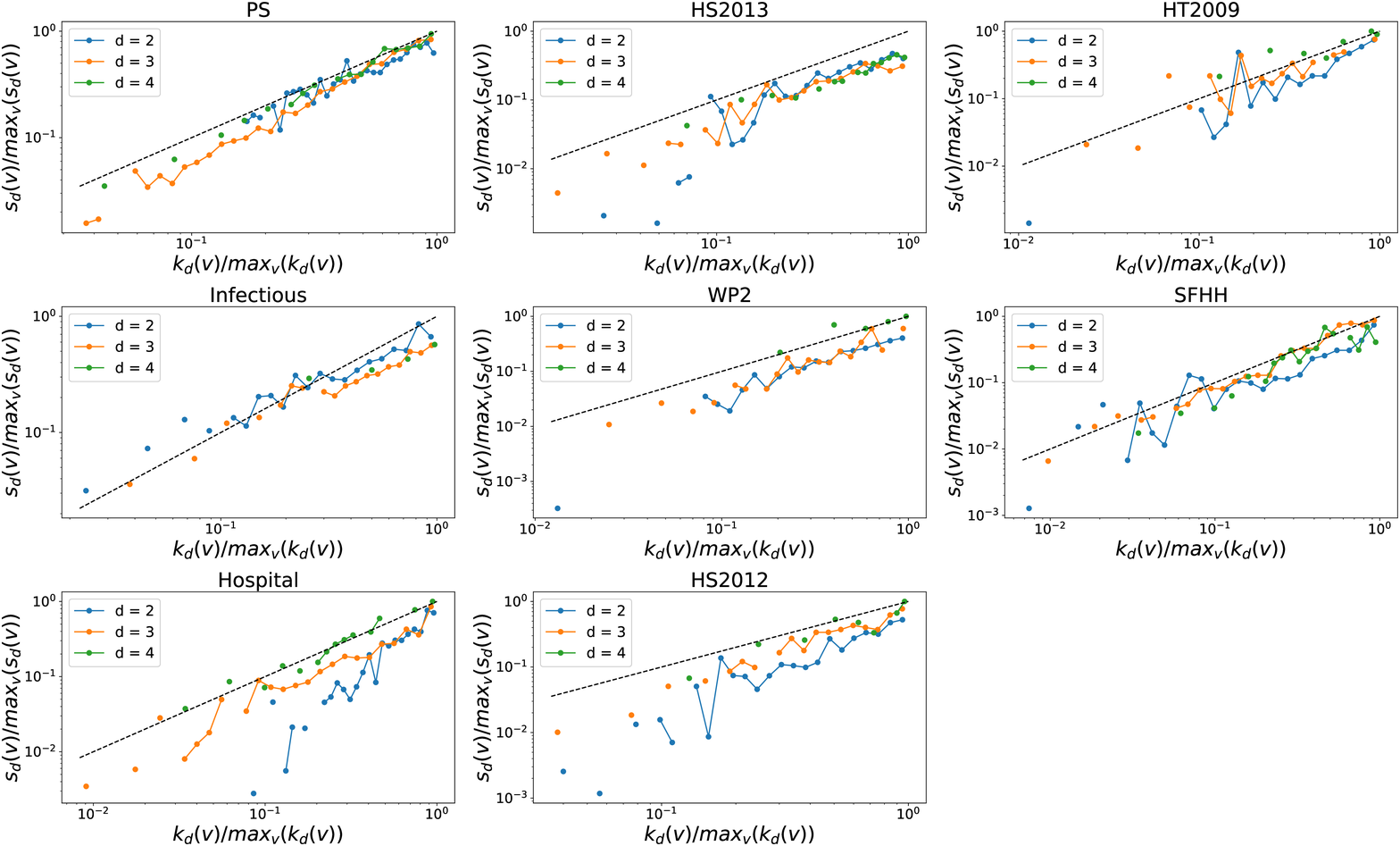}
    \caption{The $d$-strength $s_{d}(v)$ versus the the $d$-degree $k_{d}(v)$ of a node $v$ when $d$ is equal to 2 (blue line), 3 (yellow line) and 4 (green line) respectively in each physical contact network. Each axis (e.g., $k_{d}(v)$) has been normalized by its maximum (e.g., $max_{v}(k_{d}(v))$). The dashed line represent the reference case $\frac{s_{d}(v)}{max_{v}(s_{d}(v))} = \frac{k_{d}(v)}{max_{v}(k_{d}(v))}$,  where $d$-strength is a linear function of the $d$-degree of nodes. Note that both axes are presented in logarithmic scales. In total 30 logarithmic bins are split for horizontal axis.}
    \label{fig:node_deg_str_corr_phys}
\end{figure}

\begin{figure}[H]
    \centering
    \includegraphics[width =0.95 \textwidth]{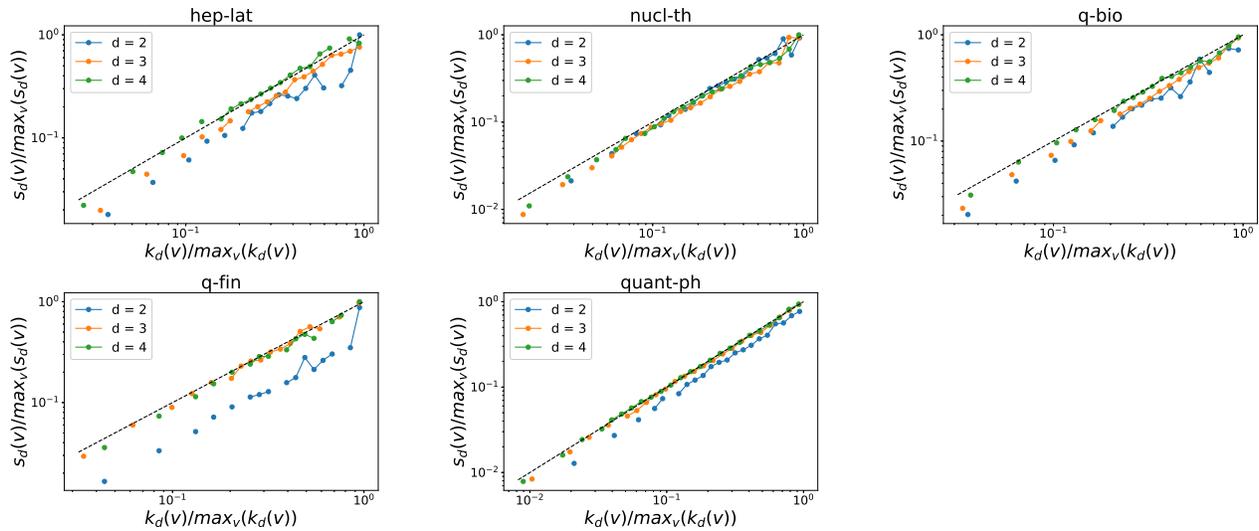}
    \caption{Average $d$-strength as function of the $d$-degree  of nodes for orders $d = 1$ (blue line), $d = 2$ (yellow line) and $d = 3$ (green line) for collaboration datasets. The dashed line represent the reference case $\frac{s_{d}(v)}{max_{v}(s_{d}(v))} = \frac{k_{d}(v)}{max_{v}(k_{d}(v))}$,  where $d$-strength is a linear function of the $d$-degree of nodes. Note that both axes are presented in logarithmic scales. In total 30 logarithmic bins are split for horizontal axis.}
    \label{fig:node_deg_str_corr_collab}
\end{figure}

The positive correlation both in the degree of a node between two different orders and in the strength of a node between two different orders can be partially explained by the high correlation between the $d$-strength and $d$-degree of a node, in every dataset as shown in Figures \ref{fig:node_deg_str_corr_phys} and \ref{fig:node_deg_str_corr_collab}.
We further observe that the $d$-strength of a node is approximately a linear function of the $d$-degree of the node at each order. This linear function means that the average number of times a node interacts with an order $d$ group (the ratio of the $d$-strength to the $d$-degree of the node) is a constant, independent of the number of distinct order $d$ groups the node interacts with. Thus, engaging in more groups of a given order $d$ will not affect an individual's average number of interactions per group.  


\subsection{Temporal correlation of events at a local egonetwork}
\label{subsec:loc_temp}
\begin{figure}[H]
    \centering
    \includegraphics[width = \textwidth]{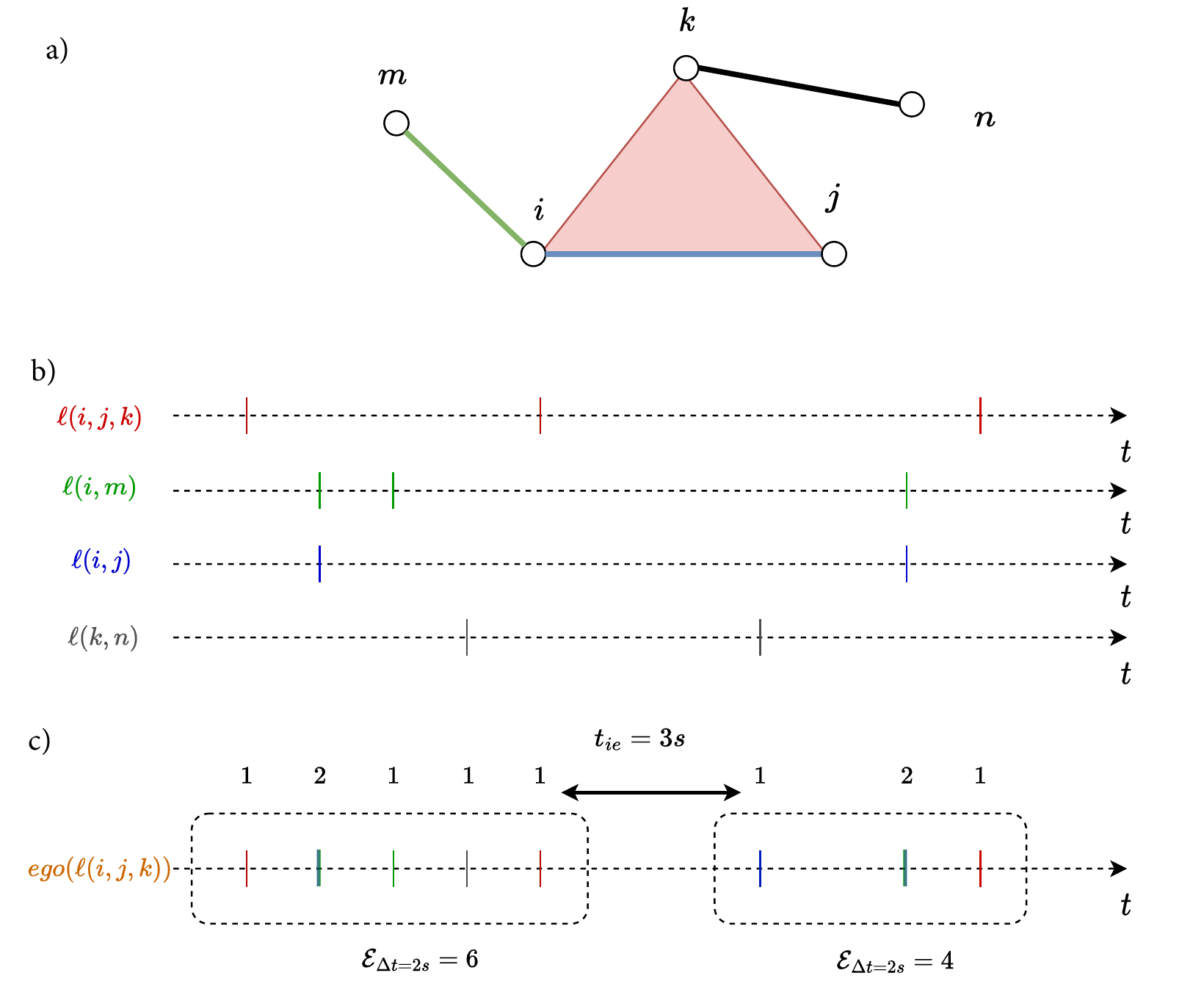}
    \caption{Schematic representation of  a) the egonetwork of the hyperlink $h(i,j,k) $, i.e. $ego(h(i,j,k))$, b) the time series associated to links belonging to $ego(h(i,j,k))$ , c) the time series of the activity of  $ego(h(i,j,k))$ , which is the sum of the time series of hyperlinks belonging to the egonetwork, and its event trains identified when  $\Delta t = 2s$.}
    \label{fig:ego_high}
\end{figure}

Since higher-order events overlap in topology, e.g., the component nodes of a higher-order event tend to participate in events of a lower order, we explore further the temporal correlation of events that occur locally in topology. 
The topological neighborhood of a hyperlink $h_d$ of order $d$, so called the egonetwork $ego(h_d)$ centered at $h_d$, is defined as the union of the hyperlink $h_d$ and all hyperlinks with an order lower than $d$ that share at least one node with $h_d$ in the higher-order aggregated network.
We construct the time series of the aggregated activity of an egonetwork $ego(h_d)$, as the sum of the time series of hyperlinks belonging to $ego(h_d)$, as shown in Figure \ref{fig:ego_high}.
We then evaluate the temporal correlation of the time series of an egonetwork $ego(h_d)$, to understand whether the activation of the center hyperlink $h_d$ tend to cluster in time with the activation of the other low order hyperlinks in the egonetwork $ego(h_d)$.

Our analysis method is based on the concept of event trains, proposed by Karsai et al. \cite{karsai2012universal}. A train of events is a sequence of consecutive events whose inter-event times are shorter than or equal to a reference temporal interval $\Delta t$ and separated from the other contacts by an inter-event times larger than $\Delta t$. Given a $\Delta t$ and an activity time series of an egonetwork $ego(h_d)$, trains can be identified, as exemplified in Figure \ref{fig:ego_high}. Given $\Delta t$ and an order $d$, we identify all the trains for each activity series of the egonetwork centered at each order $d$ hyperlink. The size of a train is the number of events the train contains. Then, we examine the size distribution $Pr[\mathcal{S}^*_{d} = s]$ of the identified trains in which a center hyperlink has been activated at least once. 
The timescales of physical contacts and collaboration networks are different. The two classes are measured per step of seconds and day respectively.  To illustrate our method and findings we consider $\Delta t = 60 s$ ($ 60 d$) in physical contact (collaboration) networks to identify the trains in each egonetwork. The choice $\Delta t = 60 s$ is also motivated by the observation in Figure \ref{fig:temp_top_cross} that we start to observe the positive temporal and topological correlation of higher-order events since $\Delta t $ is about $ 80 s$ in physical contact networks. Moreover, we observe the same when $\Delta t = 120 s$ ($120 d$) in physical contact (collaboration) networks in the coming analysis.

\begin{figure}[!h]
    \centering
    \includegraphics[width = \textwidth]{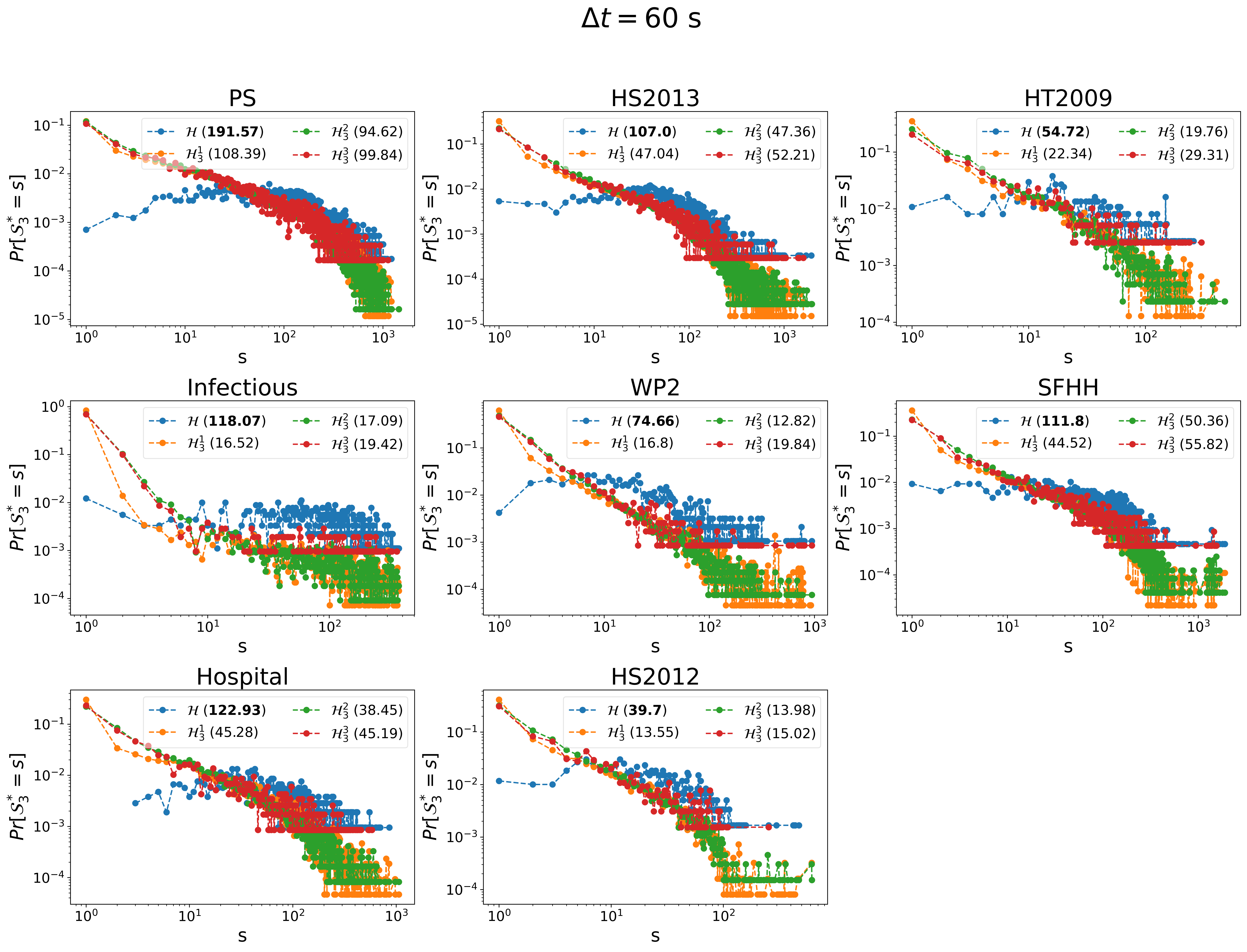}
    \caption{Probability distribution $Pr[\mathcal{S}^*_{3}=s]$ of the size $\mathcal{S}^*_{3}$ of trains (obtained from the activity series of egonetworks centered at each order 3 hyperlink), where a center link is activated at least once, in each physical contact network $\mathcal{H}$ (blue) and its three randomized reference models $\mathcal{H}^1_3$ (yellow), $\mathcal{H}^2_3$ (green) and $\mathcal{H}^3_3$ (red). To identify the trains, we consider $\Delta t = 60s$. For each network, the average size of the trains is reported. The maximum average size among network $\mathcal{H}$, $\mathcal{H}^1_3$, $\mathcal{H}^2_3$ and $\mathcal{H}^3_3$ is in bold. The horizontal and vertical axes are presented in logarithmic scale.}
    \label{fig:ego_train60}
\end{figure}

\begin{figure}[!h]
    \centering
    \includegraphics[width = \textwidth]{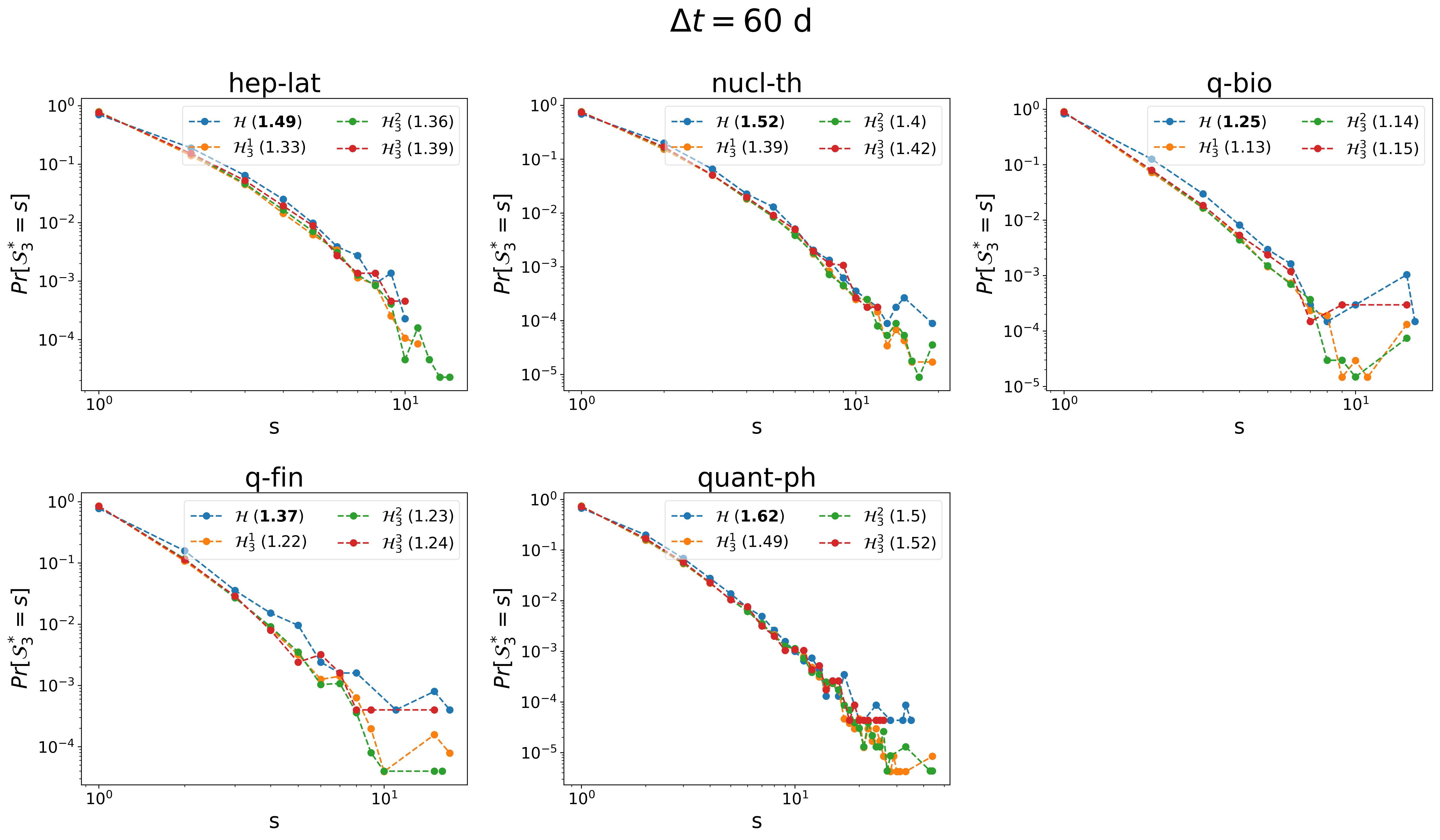}
    \caption{Probability distribution $Pr[\mathcal{S}^*_{3}=s]$ of the size $\mathcal{S}^*_{3}$ of trains (obtained from the activity series of egonetworks centered at each order 3 hyperlink), where a center link is activated at least once, in each collaboration network $\mathcal{H}$ (blue) and its three randomized reference models $\mathcal{H}^1_3$ (yellow), $\mathcal{H}^2_3$ (green) and $\mathcal{H}^3_3$ (red). To identify the trains, we consider $\Delta t = 60s$. For each network, the average size of the trains is reported. The maximum average size among network $\mathcal{H}$, $\mathcal{H}^1_3$, $\mathcal{H}^2_3$ and $\mathcal{H}^3_3$ is in bold. The horizontal and vertical axes are presented in logarithmic scale.}
    \label{fig:ego_train60_collab}
\end{figure}

Figure \ref{fig:ego_train60} and \ref{fig:ego_train60_collab} show the train size distribution $Pr[\mathcal{S}^*_{3} = s]$ of the egonetworks centered at each order $3$ hyperlink in each physical and collaboration network $\mathcal{H}$ and its three null models $\mathcal{H}^1_{3}$, $\mathcal{H}^2_{3}$, $\mathcal{H}^3_{3}$. The three randomized reference models systematically destroy temporal properties of order 3 events only, while preserving properties of events of any other order. In physical contact networks, the train size is evidently larger on average than that in their corresponding randomized networks. This indicates that an order $3$ event tend to occur close in time with many local order $2$ events, forming large trains. The trains in collaboration networks are, however, not evidently longer than those in randomized reference models on average. We found similar when considering $\Delta t = 120s$ for physical contacts and $\Delta t = 120d$ for collaboration networks (see Figures \ref{fig:ego_train120} and \ref{fig:ego_train120_coll} in Supplementary Material). 

The temporal correlation analysis of local events helps explain the interrelation of topological and temporal distance of higher-order events discovered in Subsection \ref{subsec:topo_tempo_interr}. In physical contact (collaboration) networks, we observe evident (no evident) correlation between topological and temporal distance of events with different orders. Consistently, whereas events overlap in component nodes in both types of networks, local events, thus events close in topology are strongly (weakly or not) correlated in time, in forming long trains, in physical contact (collaboration) networks. 
In networks where the interrelation between topological and temporal distance of events is more evident (e.g., Infectious, WP2 and Hospital), the correlation of local events in time also tends to be stronger (average train size observed in real-work network is evidently larger than that of randomized reference models).

The detected differences between physical contact and collaboration networks may be explained by the fact that physical interactions are driven by physical proximity. For example, individuals that have a group interaction are close in physical distance, which may facility the interaction of a subgroup, resulting in events close in time and topology. 

\section{Conclusion}

In this paper, we have proposed a method to systematically characterize temporal and topological properties of events of arbitrary orders.
We applied our methods to 8 physical contact and 5 collaboration higher-order evolving networks and observe their difference.
 In physical contacts, events close in time tend to occur also close in topology. Moreover, events usually overlap in component nodes and these local events overlapping in component nodes are also usually correlated in time. Such temporal correlation of local events supports again the correlation between temporal and topological distances of events observed in our first analysis. Differently, in collaboration networks, the temporal and topological correlation of events is either weak or absent. Despite events also overlap in component nodes, their temporal correlation almost disappears in collaboration networks. The detected dissimilarities between physical contacts and collaboration networks could be related to a fundamental difference between the two kind of networks. In physical contacts individuals participate in events driven by physical proximity. The physical proximity of individuals that participate in a higher-order event may facilitate interaction of them or a subgroup in the near future. The time of scientific collaborations are likely driven more by their content and creation process.  
 
Via our analysis of the topological overlap of events with different orders in component nodes, we also observe similarities between the two kinds of networks. 
Nodes that participate in many events (groups) of a given order tend to interact in many events (groups) of a different order. Hence, nodes are  consistent in interactions with respect to frequency and diversity across different orders.

Our method explores the temporal and topological relation of the basic building block of events, the activations of fully connected cliques. A promising direction could be generalizing this method to the activations of relevant motifs, and to investigate the interplay between topological location and temporal delay of such structures. Beyond, our method can be applied to compare different classes of networks (e.g. biological, brain or collaboration networks) and to explore how detected properties/patterns of a network can influence the dynamic  processes unfolding on the network. Finally, the topological and temporal properties of events detected in this paper could foster higher-order evolving network models that better reproduce patterns observed so far.

\bibliography{main}
\section*{Funding}
This work is supported by Netherlands Organisation for Scientific Research NWO (TOP Grant no. 612.001.802).
\newpage
\setcounter{figure}{0}
\renewcommand{\figurename}{Fig.}
\renewcommand{\thefigure}{S\arabic{figure}}
\appendix
\section*{Supplementary Material}
\subsection*{General Statistics}
\begin{figure}[H]
    \centering
    \includegraphics[width = 0.81\textwidth]{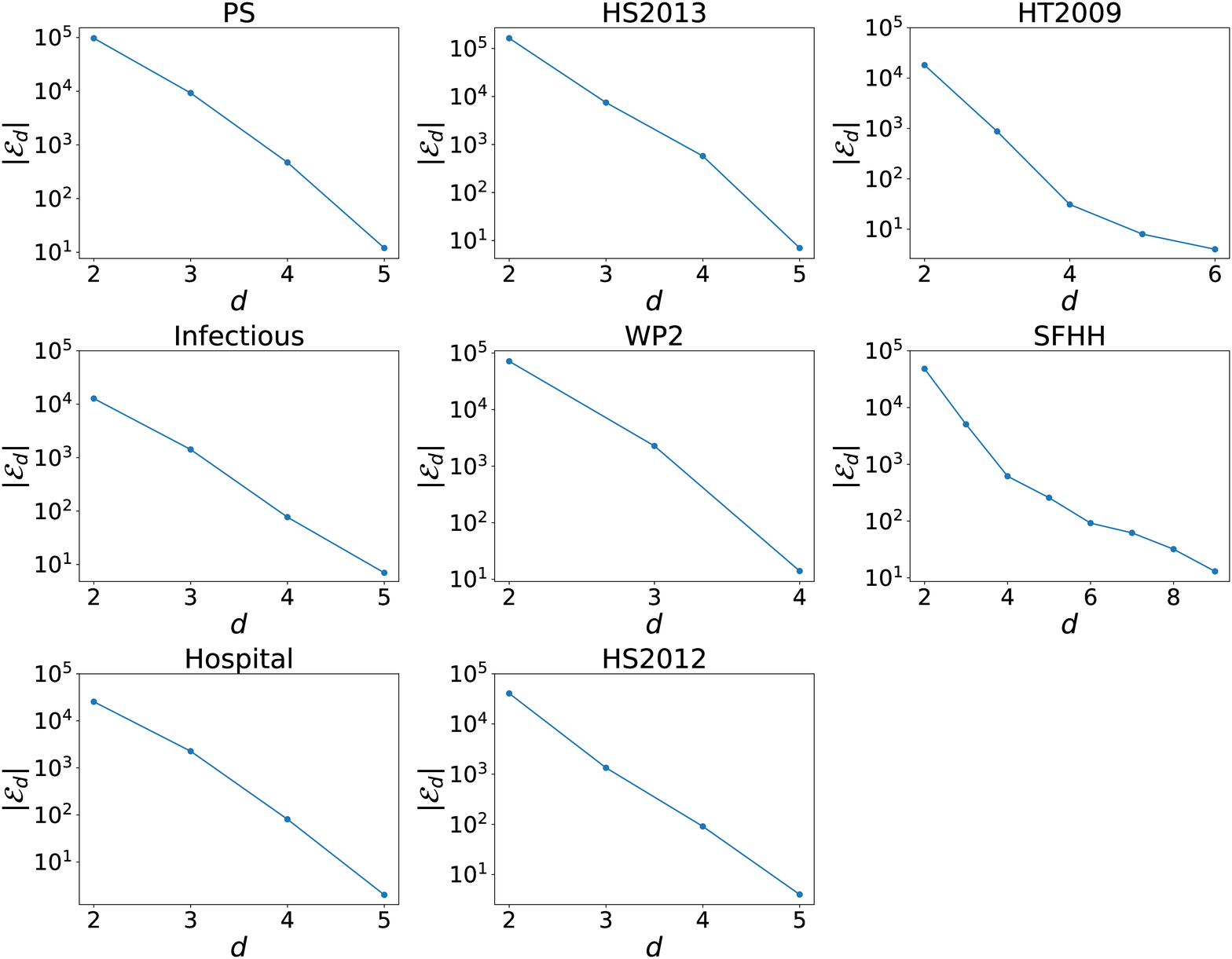}
    \caption{Total number of events ($|\mathcal{E}_d|$) for each order $d$ in physical contact networks. Vertical axis is presented in logarithmic scale.}
    \label{fig:hypercount_phys}
\end{figure}

\begin{figure}[H]
    \centering
    \includegraphics[width = 0.81\textwidth]{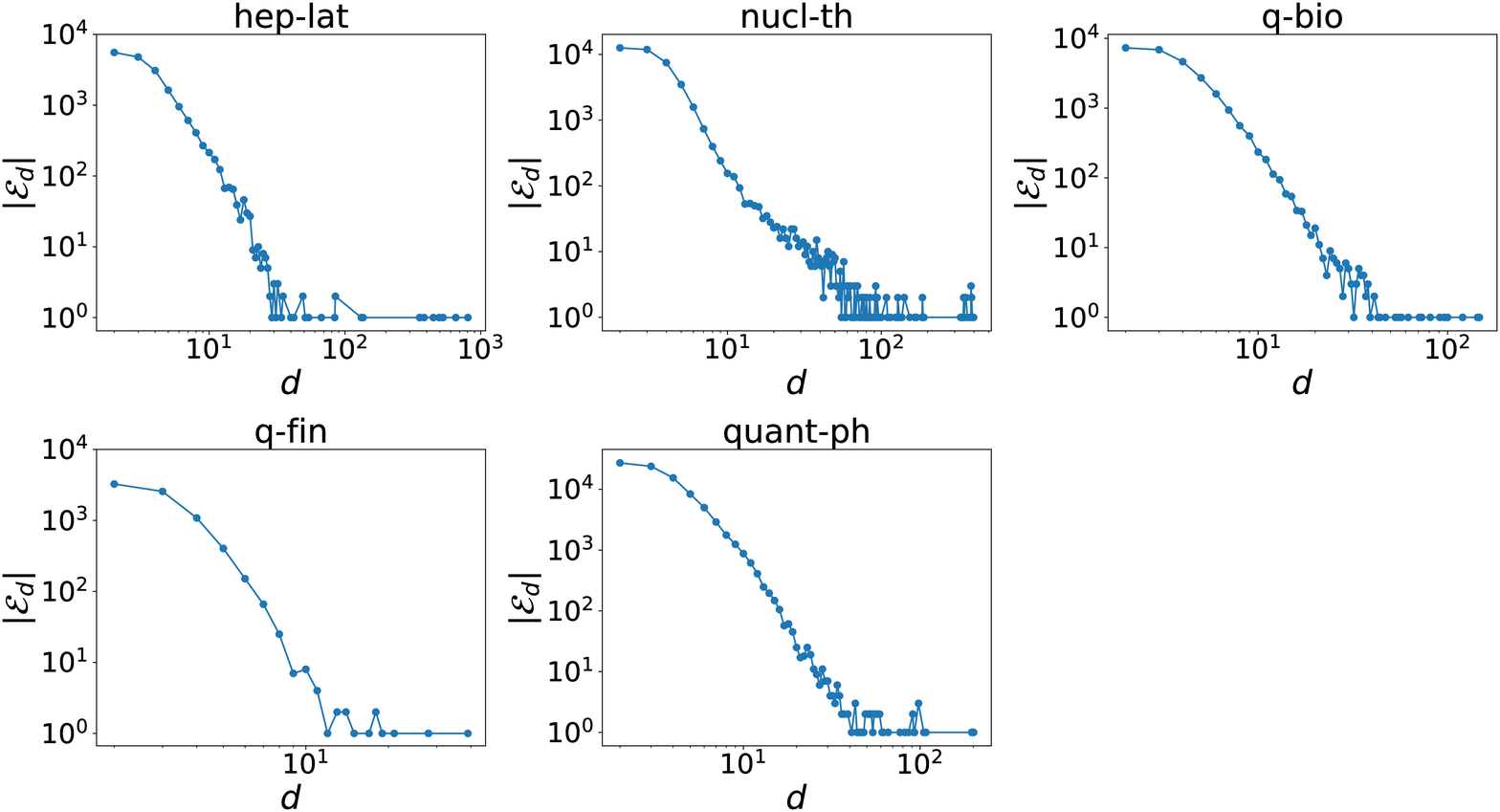}
    \caption{Total number of events ($|\mathcal{E}_d|$) for each order $d$ in collaboration networks. Vertical axis is presented in logarithmic scale.}
    \label{fig:hypercount_collab}
\end{figure}

\begin{figure}[H]
    \centering
    \includegraphics[width = 0.81\textwidth]{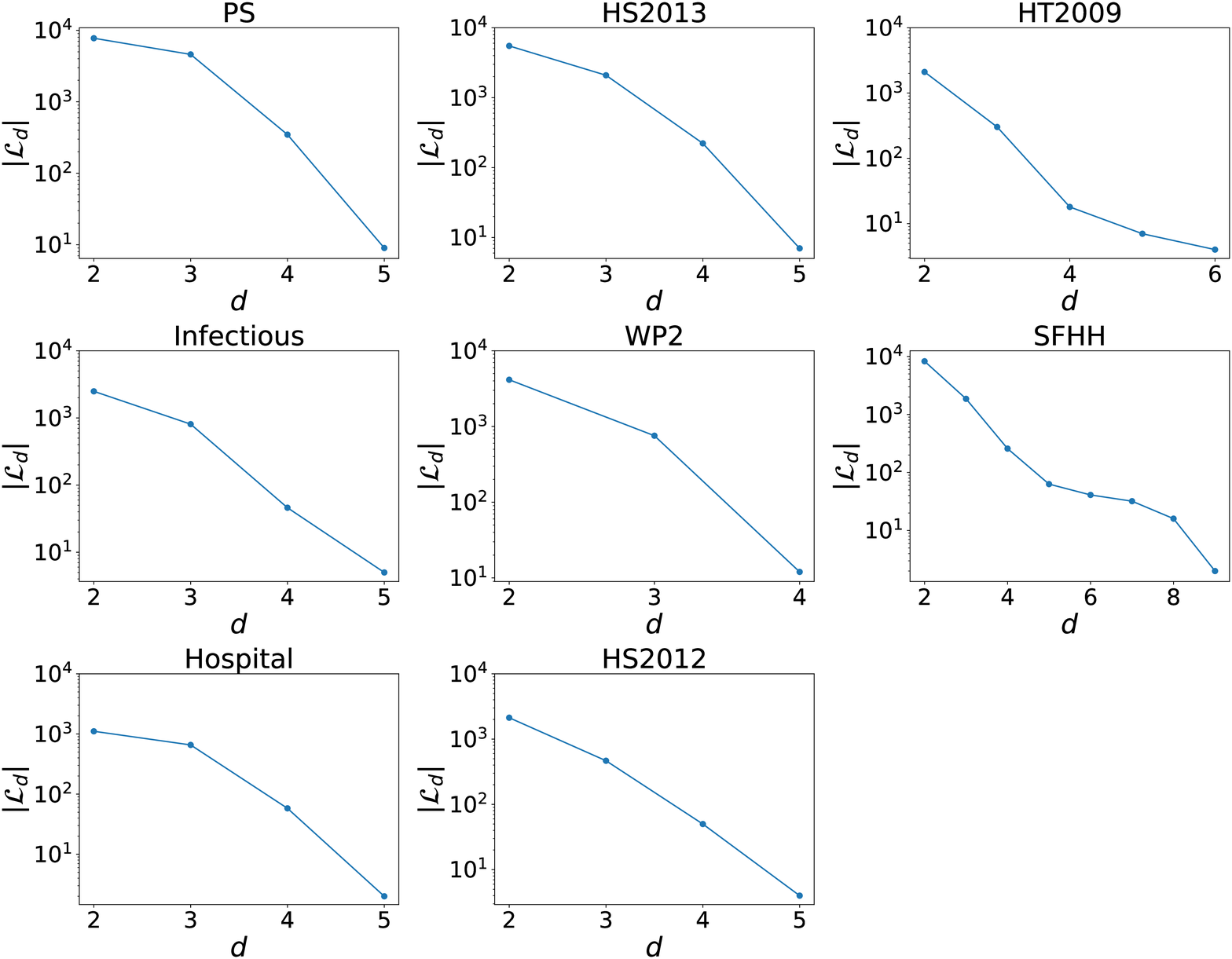}
    \caption{Total number of hyperlinks ($|\mathcal{L}_d|$) in the time aggregated higher-order network for each order $d$ for physical contact datasets. Vertical axis is presented in logarithmic scale.}
    \label{fig:active_hyper}
\end{figure}

\begin{figure}[H]
    \centering
    \includegraphics[width = 0.81\textwidth]{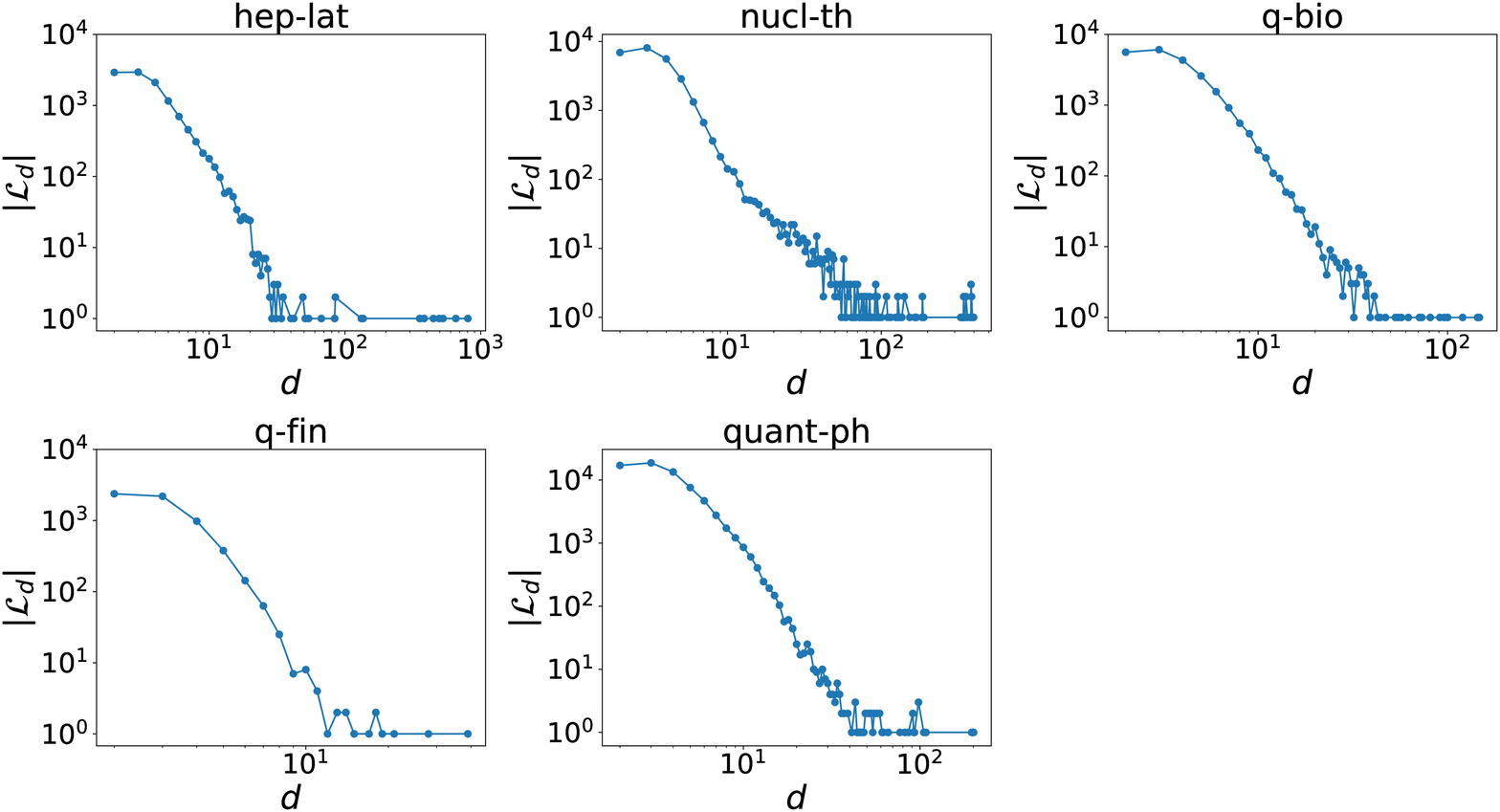}
    \caption{Total number of hyperlinks ($|\mathcal{L}_d|$) for each order $d$ in collaboration networks. Vertical axis is presented in logarithmic scale.}
    \label{fig:active_hyper_collab}
\end{figure}

\subsection*{Temporal-topological correlation of events}
\begin{figure}[!h]
    \centering
    \includegraphics[width = \textwidth]{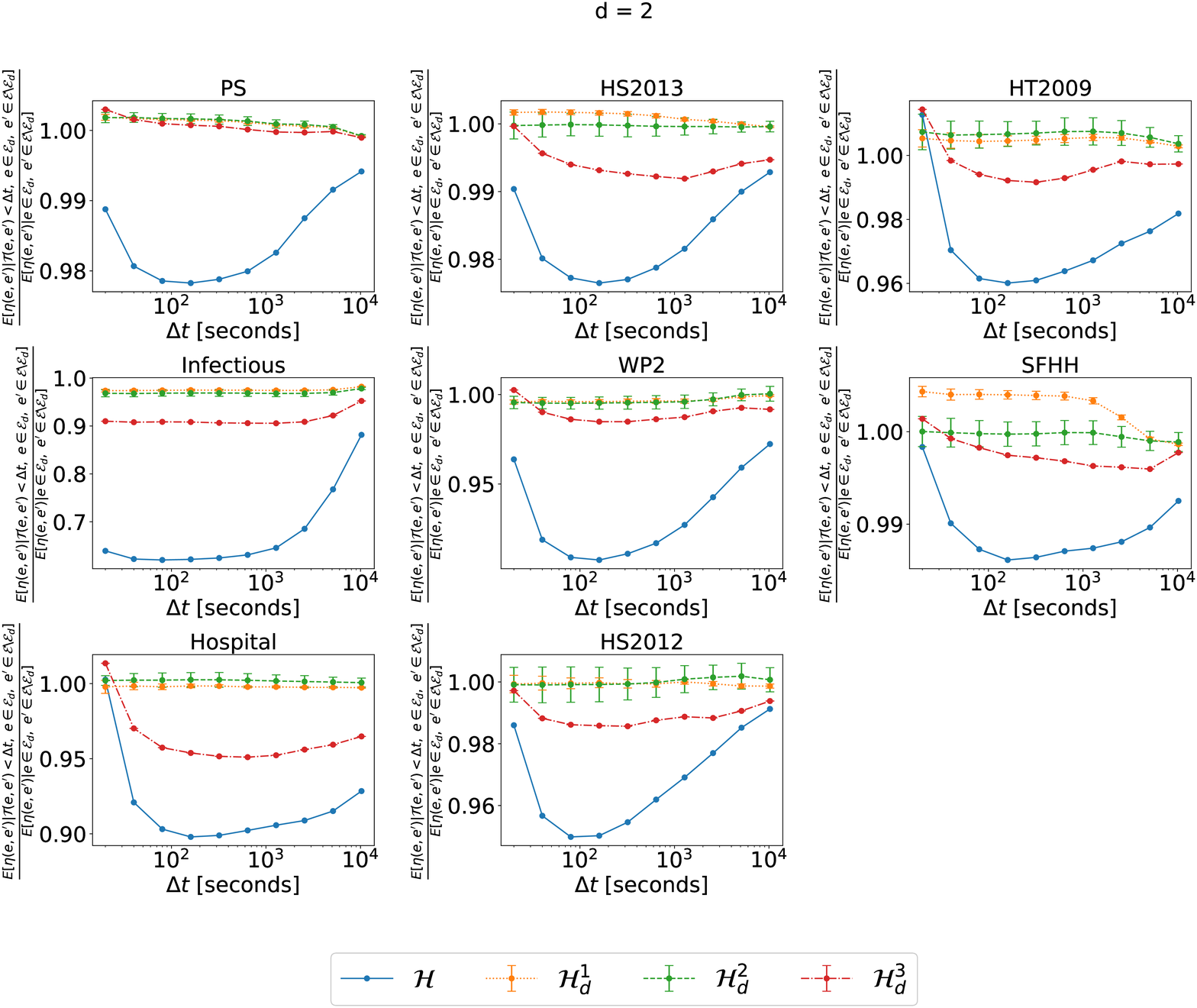}
    \caption{The normalized average topological distance $\frac{E[\eta(e,e') | \mathcal{T} (e,e') < \Delta t,\ e\in \mathcal{E}_d,\ e' \in \mathcal{E}\setminus \mathcal{E}_d ]}{E[\eta(e,e')|\ e\in \mathcal{E}_d,\ e' \in \mathcal{E}\setminus \mathcal{E}_d]}$, between an order $d=2$ event and an event of a different order, in each physical contact network and its corresponding three randomized null models $\mathcal{H}^1_d$ (yellow), $\mathcal{H}^2_d$ (green) and $\mathcal{H}^3_d$ (red), which preserve or destroy specific properties of order $d=3$ events. $\lim_{\Delta t\to\infty} E[\eta(e,e') | \mathcal{T} (e,e') < \Delta t,\ e\in \mathcal{E}_d,\ e' \in \mathcal{E}\setminus \mathcal{E}_d ] =E[\eta(e,e')|\ e\in \mathcal{E}_d,\ e' \in \mathcal{E}\setminus \mathcal{E}_d]$ for any $d$. The horizontal axes are presented in logarithmic scale.
    For each dataset, the results of the three corresponding randomized models are obtained from 10 independent realizations.}
    \label{fig:temp_top2_cross}
\end{figure}

\begin{figure}[!h]
    \centering
    \includegraphics[width = \textwidth]{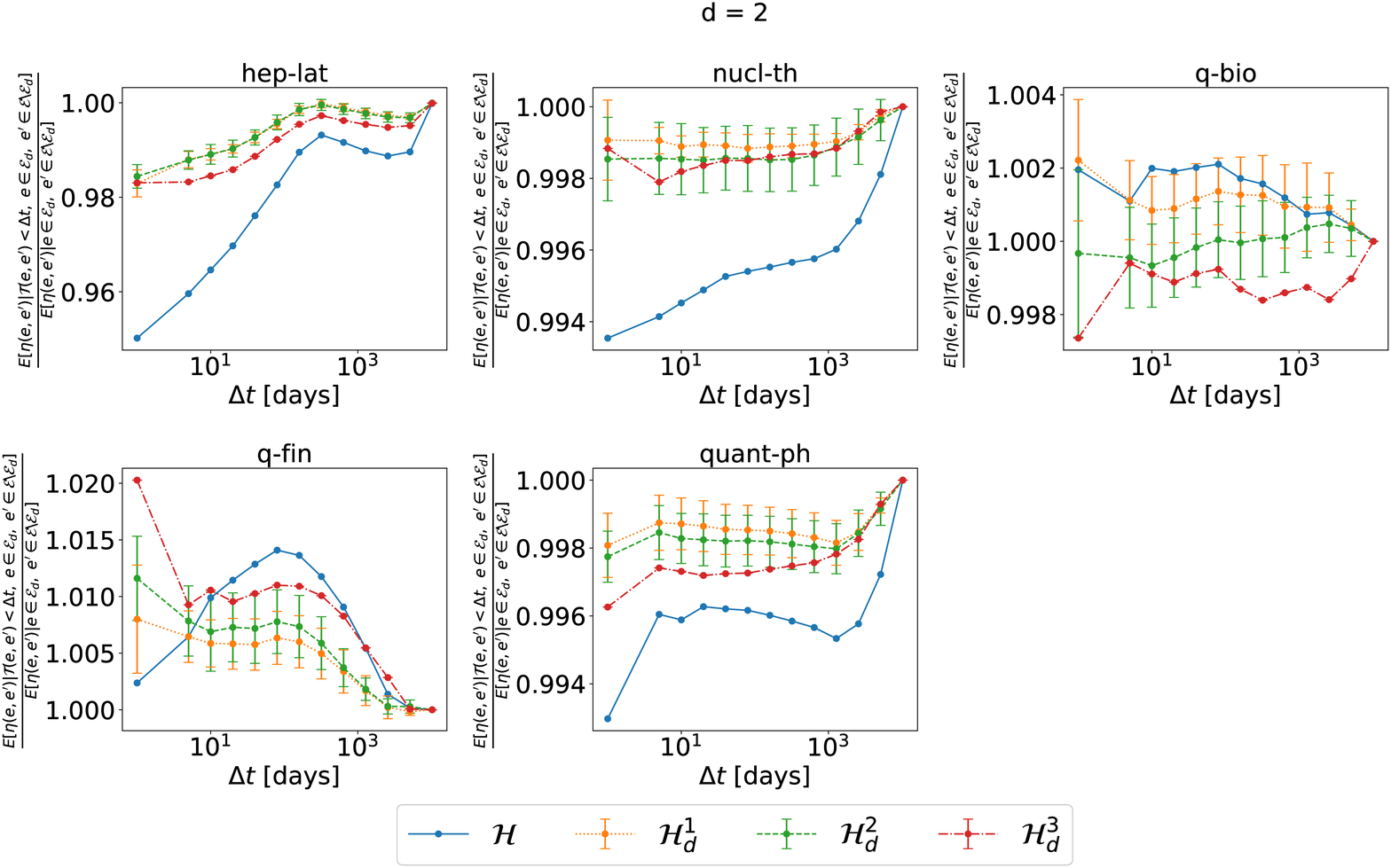}
    \caption{The normalized average topological distance $\frac{E[\eta(e,e') | \mathcal{T} (e,e') < \Delta t,\ e\in \mathcal{E}_d,\ e' \in \mathcal{E}\setminus \mathcal{E}_d ]}{E[\eta(e,e')|\ e\in \mathcal{E}_d,\ e' \in \mathcal{E}\setminus \mathcal{E}_d]}$, between an order $d=2$ event and an event of a different order, in each collaboration network and its corresponding three randomized null models $\mathcal{H}^1_d$ (yellow), $\mathcal{H}^2_d$ (green) and $\mathcal{H}^3_d$ (red), which preserve or destroy specific properties of order $d=2$ events. $\lim_{\Delta t\to\infty} E[\eta(e,e') | \mathcal{T} (e,e') < \Delta t,\ e\in \mathcal{E}_d,\ e' \in \mathcal{E}\setminus \mathcal{E}_d ] =E[\eta(e,e')|\ e\in \mathcal{E}_d,\ e' \in \mathcal{E}\setminus \mathcal{E}_d]$ for any $d$. The horizontal axes are presented in logarithmic scale.
    For each dataset, the results of the three corresponding randomized models are obtained from 10 independent realizations.}
    \label{fig:temp_top2_cross_collab}
\end{figure}

\begin{figure}[!h]
    \centering
    \includegraphics[width = \textwidth]{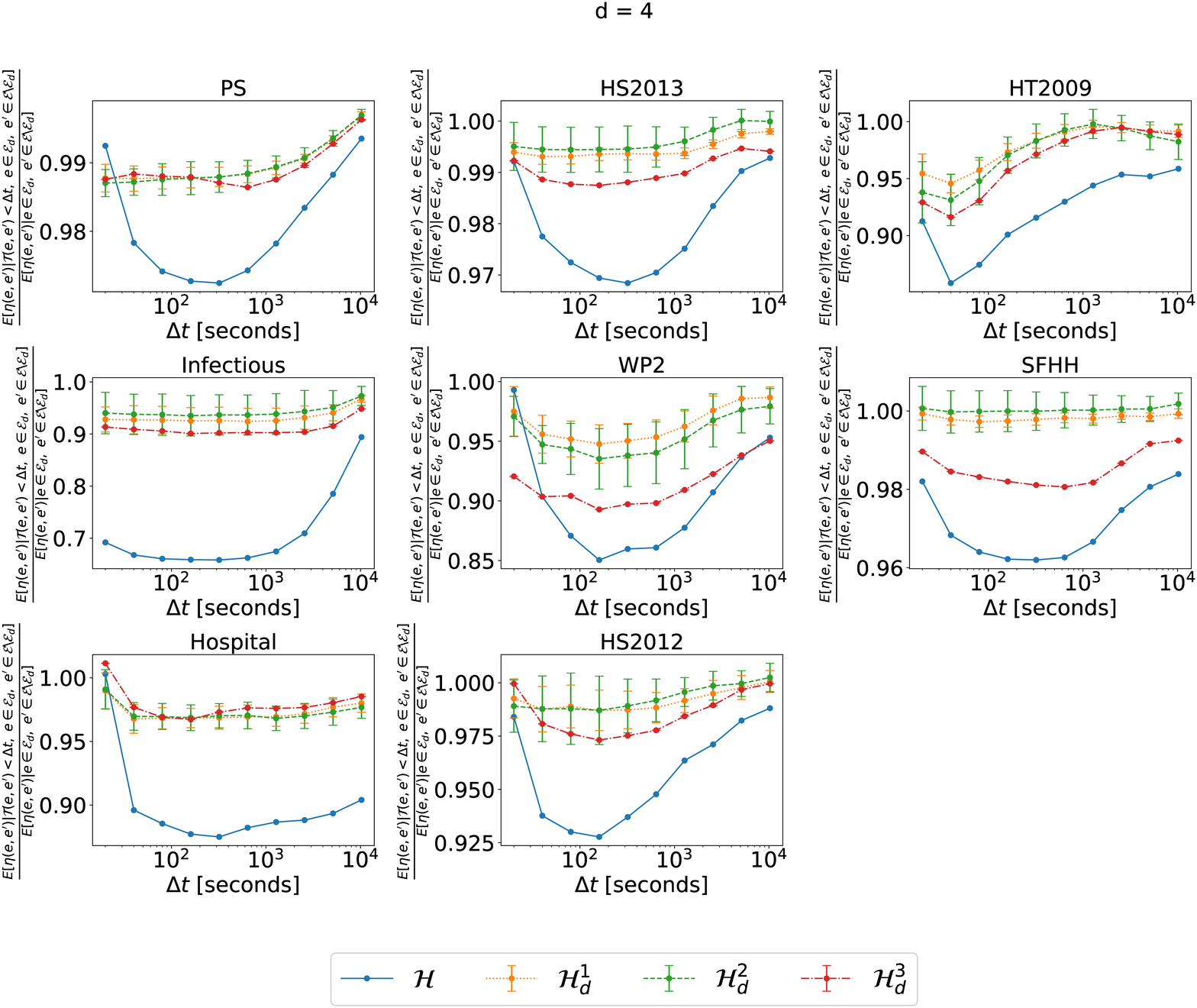}
    \caption{The normalized average topological distance $\frac{E[\eta(e,e') | \mathcal{T} (e,e') < \Delta t,\ e\in \mathcal{E}_d,\ e' \in \mathcal{E}\setminus \mathcal{E}_d ]}{E[\eta(e,e')|\ e\in \mathcal{E}_d,\ e' \in \mathcal{E}\setminus \mathcal{E}_d]}$, between an order $d=4$ event and an event of a different order, in each physical contact network and its corresponding three randomized null models $\mathcal{H}^1_d$ (yellow), $\mathcal{H}^2_d$ (green) and $\mathcal{H}^3_d$ (red), which preserve or destroy specific properties of order $d=4$ events. $\lim_{\Delta t\to\infty} E[\eta(e,e') | \mathcal{T} (e,e') < \Delta t,\ e\in \mathcal{E}_d,\ e' \in \mathcal{E}\setminus \mathcal{E}_d ] =E[\eta(e,e')|\ e\in \mathcal{E}_d,\ e' \in \mathcal{E}\setminus \mathcal{E}_d]$ for any $d$. The horizontal axes are presented in logarithmic scale.
    For each dataset, the results of the three corresponding randomized models are obtained from 10 independent realizations.}
    \label{fig:temp_top3_cross}
    \end{figure}
    
\begin{figure}[!h]
    \centering
    \includegraphics[width = \textwidth]{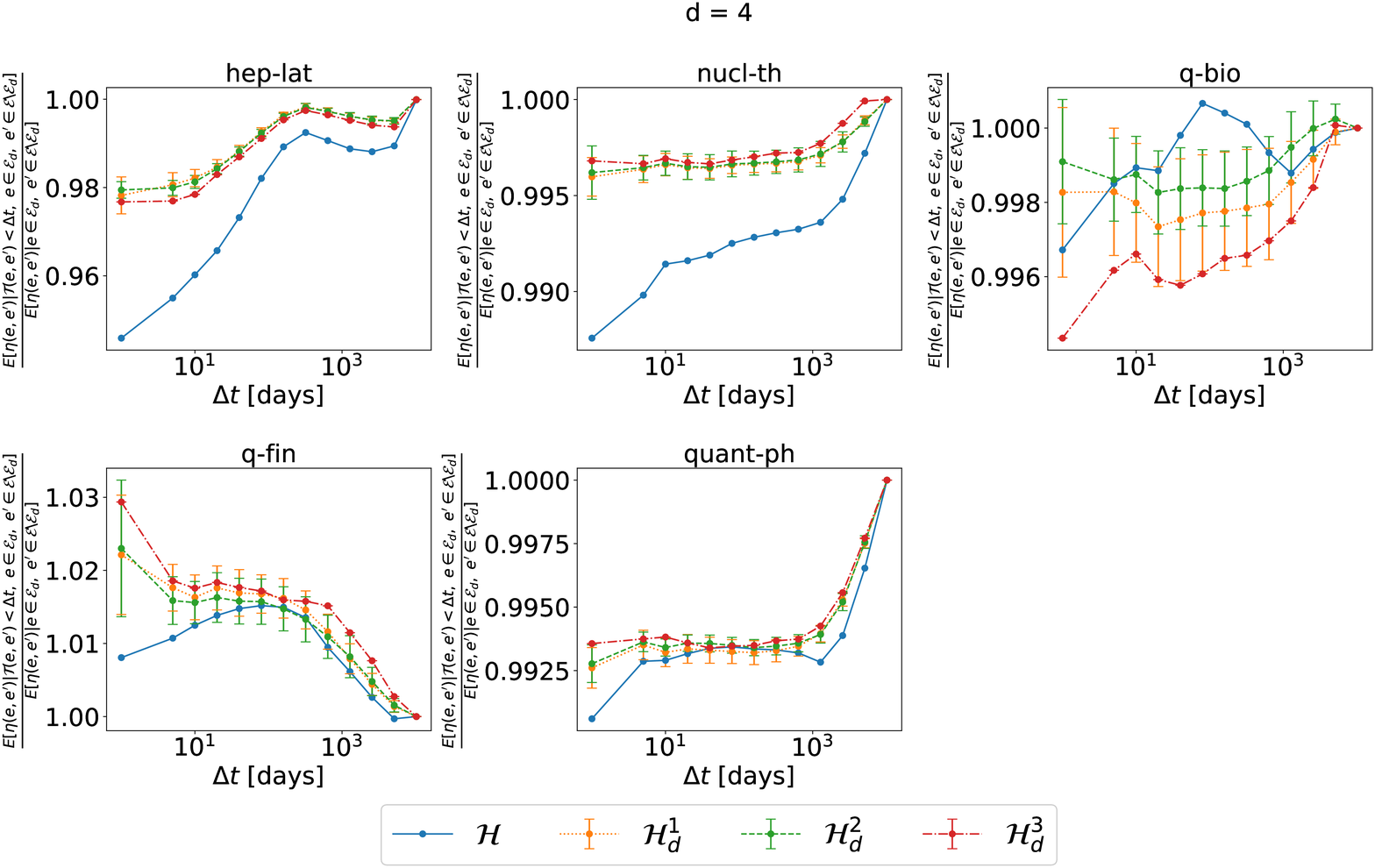}
    \caption{The normalized average topological distance $\frac{E[\eta(e,e') | \mathcal{T} (e,e') < \Delta t,\ e\in \mathcal{E}_d,\ e' \in \mathcal{E}\setminus \mathcal{E}_d ]}{E[\eta(e,e')|\ e\in \mathcal{E}_d,\ e' \in \mathcal{E}\setminus \mathcal{E}_d]}$, between an order $d=4$ event and an event of a different order, in each collaboration network and its corresponding three randomized null models $\mathcal{H}^1_d$ (yellow), $\mathcal{H}^2_d$ (green) and $\mathcal{H}^3_d$ (red), which preserve or destroy specific properties of order $d=3$ events. $\lim_{\Delta t\to\infty} E[\eta(e,e') | \mathcal{T} (e,e') < \Delta t,\ e\in \mathcal{E}_d,\ e' \in \mathcal{E}\setminus \mathcal{E}_d ] =E[\eta(e,e')|\ e\in \mathcal{E}_d,\ e' \in \mathcal{E}\setminus \mathcal{E}_d]$ for any $d$. The horizontal axes are presented in logarithmic scale.
    For each dataset, the results of the three corresponding randomized models are obtained from 10 independent realizations.}
    \label{fig:temp_top3_cross_collab}
    \end{figure}

\begin{figure}[!h]
    \centering
    \includegraphics[width = \textwidth]{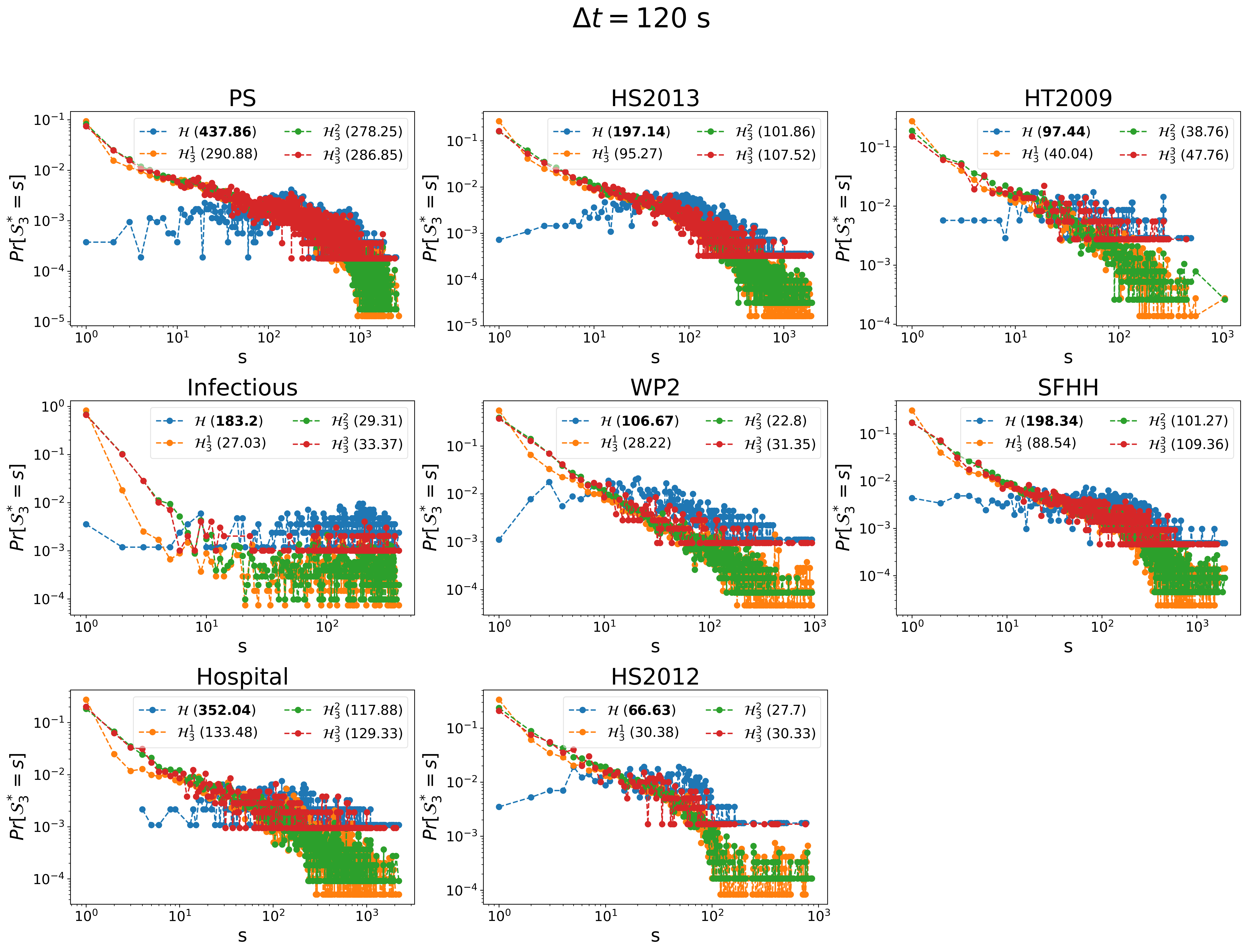}
    \caption{Probability distribution $Pr[\mathcal{S}^*_{3}=s]$ of the size $\mathcal{S}^*_{3}$ of trains (obtained from the activity series of egonetworks centered at each order 3 hyperlink), where a center link is activated at least once, in each physical contact network $\mathcal{H}$ (blue) and its three randomized reference models $\mathcal{H}^1_3$ (yellow), $\mathcal{H}^2_3$ (green) and $\mathcal{H}^3_3$ (red). To identify the trains, we consider $\Delta t = 120s$. For each network, the average size of the trains is reported. The maximum average size among network $\mathcal{H}$, $\mathcal{H}^1_3$, $\mathcal{H}^2_3$ and $\mathcal{H}^3_3$ is in bold. The horizontal and vertical axes are presented in logarithmic scale.}
    \label{fig:ego_train120}
\end{figure}

\begin{figure}[!h]
    \centering
    \includegraphics[width = \textwidth]{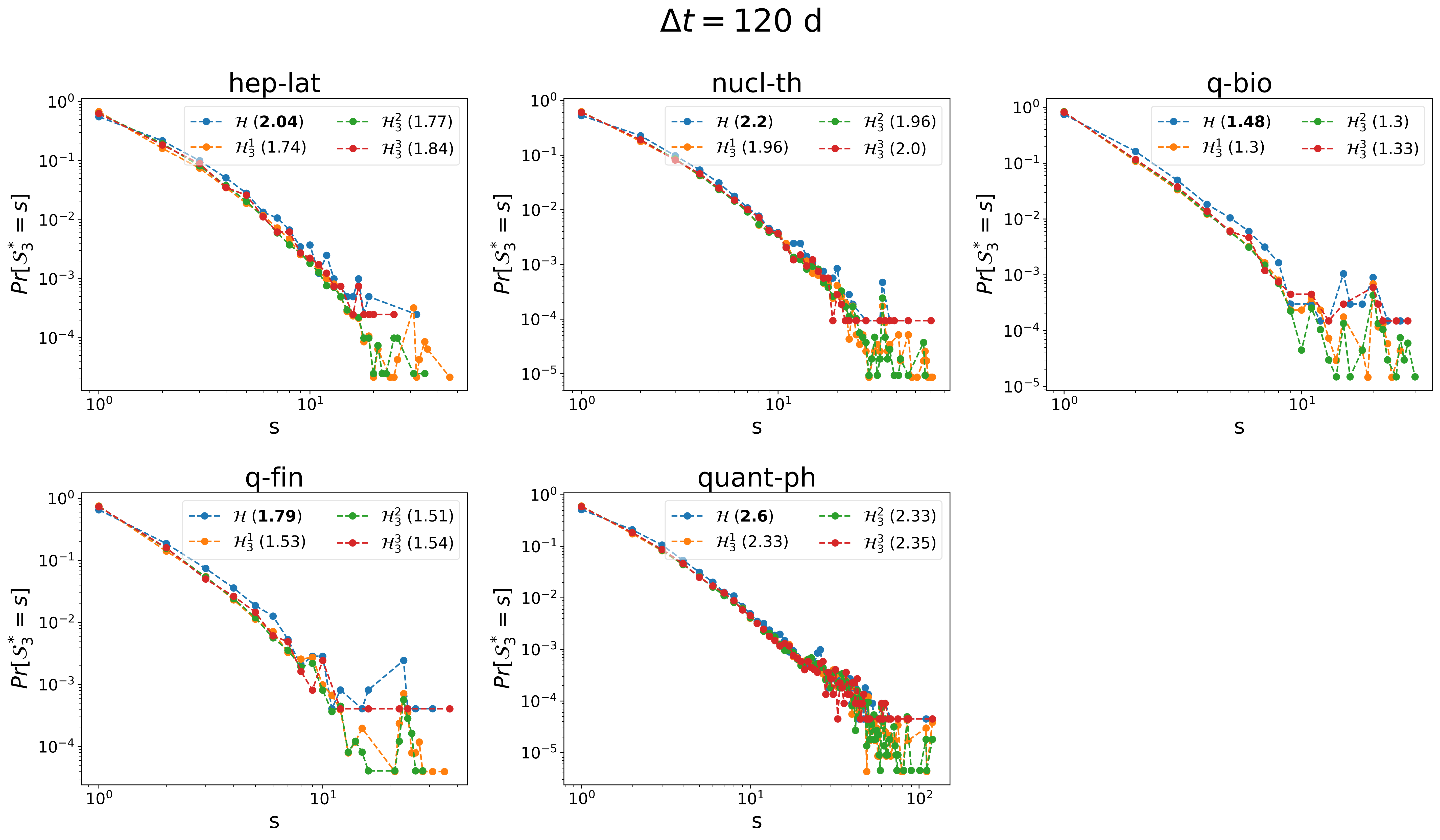}
    \caption{Probability distribution $Pr[\mathcal{S}^*_{3}=s]$ of the size $\mathcal{S}^*_{3}$ of trains (obtained from the activity series of egonetworks centered at each order 3 hyperlink), where a center link is activated at least once, in each collaboration network $\mathcal{H}$ (blue) and its three randomized reference models $\mathcal{H}^1_3$ (yellow), $\mathcal{H}^2_3$ (green) and $\mathcal{H}^3_3$ (red). To identify the trains, we consider $\Delta t = 120d$. For each network, the average size of the trains is reported. The maximum average size among network $\mathcal{H}$, $\mathcal{H}^1_3$, $\mathcal{H}^2_3$ and $\mathcal{H}^3_3$ is in bold. The horizontal and vertical axes are presented in logarithmic scale.}
    \label{fig:ego_train120_coll}
\end{figure}

\end{document}